\documentclass[10pt,journal,comsoc]{IEEEtran}
\usepackage[absolute,showboxes]{textpos}

\usepackage[T1]{fontenc}

\ifCLASSINFOpdf
\else
\fi
\usepackage{amsmath}
\usepackage[cmintegrals]{newtxmath}

\usepackage{cite}
\usepackage[para]{footmisc}
\usepackage{algorithmic}
\usepackage{enumitem}
\usepackage{graphicx}
\graphicspath{{./img/}}
\usepackage{textcomp}
\usepackage[table,xcdraw]{xcolor}
\usepackage{diagbox}
\usepackage{colortbl}
\usepackage{multirow}
\usepackage{url}
\usepackage{nohyperref} 
\usepackage[binary-units,per-mode=symbol]{siunitx}
\usepackage{subfig}
\usepackage[nohyperlinks,nolist]{acronym} 
\usepackage[]{algorithm2e}
\usepackage{tikz}
\usepackage{booktabs}
\usepackage{collcell}
\usepackage{hhline}
\usepackage{tabularx}
\newcommand*{\MinNumber}{0.0}%
\newcommand*{\MidNumber}{0.5} %
\newcommand*{\MaxNumber}{1.0}%
\newcommand{\ApplyGradient}[1]{%
        \ifdim #1 pt > \MidNumber pt
            \pgfmathsetmacro{\PercentColor}{max(min(100.0*(#1 - \MidNumber)/(\MaxNumber-\MidNumber),100.0),0.00)} %
            \hspace{-0.33em}\colorbox{green!\PercentColor!yellow}{#1}
        \else
            \pgfmathsetmacro{\PercentColor}{max(min(100.0*(\MidNumber - #1)/(\MidNumber-\MinNumber),100.0),0.00)} %
            \hspace{-0.33em}\colorbox{red!\PercentColor!yellow}{#1}
        \fi
}
\newcolumntype{Y}{>{\centering\arraybackslash}X}
\newcolumntype{L}{>{\arraybackslash}X}
\newcolumntype{R}{>{\raggedleft\arraybackslash}X}
\newcolumntype{C}[1]{>{\centering\arraybackslash}p{#1}}
\newcolumntype{G}[1]{>{\collectcell\ApplyGradient}#1<{\endcollectcell}}
\usepackage[absolute,showboxes]{textpos}
\usepackage{adjustbox}
\usetikzlibrary{patterns}
\usetikzlibrary{decorations}
\usetikzlibrary{spy}
\usepackage{pgfplots}
\usepackage{pgfplotstable}
\pgfplotsset{compat=newest}
\usepgfplotslibrary{units}
\usepgfplotslibrary{groupplots}
\makeatletter
\pgfplotsset{
  /pgfplots/ybar legend/.style={
    legend image code/.code={
      \draw [#1] (0cm,-0.1cm) rectangle (0.15cm,0.2cm);
    };
  }
}
\pgfplotsset{
  /pgfplots/xbar legend/.style={
    legend image code/.code={
      \draw [#1] (-0.1cm,-0.05cm) rectangle (0.2cm,0.1cm);
    };
  }
}
\pgfplotsset{
    minmax/.style={
        mark=|,
        error bars/.cd,
            x dir=plus,
            x explicit,
            error mark=-,
        /pgfplots/table/.cd,
            y=Time,
            x=Min,
            x error expr=\thisrow{Max}-\thisrow{Min},
    },
    avg/.style={
        mark size=2pt,
        /pgfplots/table/.cd,
            y=Time,
            x=Avg,
    }
}
\pgfplotsset{
    legend image with text/.style={
        legend image code/.code={%
            \node[anchor=center] at (0.3cm,0cm) {#1};
        }
    },
}
\pgfplotsset{
 unit code/.code 2 args=
   \begingroup
   \protected@edef\x{\endgroup\si{#2}}\x
}
\tikzset{
  every pin/.style={fill=LightCoreGray,rectangle,rounded corners=3pt,font=\footnotesize},
  small dot/.style={fill=black,circle,scale=0.3}
  }
\pgfdeclaredecoration{stars}{initial}{\state{initial}[width=6pt,next state=star1]
{
}
\state{star1}[width=4pt,next state=gap]
{
    \pgfuseplotmark{star}
  }
\state{gap}[width=4pt,next state=star1]
{
  }
\state{final}
{
    \pgfpathmoveto{\pgfpointdecoratedpathlast}
}  

}

\pgfdeclaredecoration{starstars}{initial}{\state{initial}[width=6pt,next state=star1]
{
}
\state{star1}[width=4pt,next state=star2]
{
    \pgfuseplotmark{star}
}
\state{star2}[width=4pt,next state=gap]
{
    \pgfuseplotmark{star}
  }
\state{gap}[width=4pt,next state=star1]
{
  }
\state{final}
{
    \pgfpathmoveto{\pgfpointdecoratedpathlast}
}  
}
\tikzset{
  dashstar/.style={dash pattern=on 4pt off 4pt,postaction={decorate,decoration=stars}},
  dashstarstar/.style={dash pattern=on 4pt off 8pt,postaction={decorate,decoration=starstars}},
}
\makeatother
\usetikzlibrary{matrix}

\definecolor{CoreGray}{HTML}{BFBFBF}
\definecolor{CoreBlack}{HTML}{333333}
\definecolor{CoreDarkGray}{HTML}{5F5F5F}
\definecolor{CoreBlue}{HTML}{002E7D}
\definecolor{CoreGreen}{HTML}{6AAC8E}
\definecolor{CoreRed}{HTML}{C80000}
\definecolor{CoreYellow}{HTML}{E6AC00}
\definecolor{CoreWhite}{HTML}{FFFFFF}
\definecolor{CoreMagenta}{HTML}{7030A0}
\colorlet{LightCoreGray}{CoreGray!30}
\colorlet{LightCoreBlack}{CoreBlack!20}
\colorlet{LightCoreBlue}{CoreBlue!20}
\colorlet{LightCoreGreen}{CoreGreen!30}
\colorlet{LightCoreRed}{CoreRed!20}
\colorlet{LightCoreYellow}{CoreYellow!20}
\colorlet{LightCoreWhite}{CoreWhite!20}



\begin{document}

\bstctlcite{my:BSTcontrol}

%
\title{Secure Time-Sensitive Software-Defined Networking in Vehicles}

%
%
%

\author{
  Timo~H\"ackel, Philipp~Meyer, Franz~Korf, and~Thomas~C.~Schmidt\vspace{-13pt}
  \thanks{
    © 2022 IEEE. Personal use of this material is permitted. Permission
    from IEEE must be obtained for all other uses, in any current or future
    media, including reprinting/republishing this material for advertising or
    promotional purposes, creating new collective works, for resale or
    redistribution to servers or lists, or reuse of any copyrighted
    component of this work in other works.
    }
  \thanks{The authors are with the Department of Computer Science, Hamburg University of Applied Sciences (HAW), 20099 Hamburg, Germany (e-mail: \{\href{mailto:timo.haeckel@haw-hamburg.de}{timo.haeckel}, \href{mailto:philipp.meyer@haw-hamburg.de}{philipp.meyer}, \href{mailto:franz.korf@haw-hamburg.de}{franz.korf}, \href{mailto:t.schmidt@haw-hamburg.de}{t.schmidt}\}@haw-hamburg.de).}
  \thanks{\\\hspace*{1pt} This work is funded by the Federal Ministry of Education and Research of Germany (BMBF) within the SecVI project.}
}

%
%

\markboth{IEEE Transactions on Vehicular Technology}%
{H\"ackel \MakeLowercase{\textit{et al.}}: Secure TSSDN in Vehicles}
%



\maketitle

\setlength{\TPHorizModule}{\paperwidth}
\setlength{\TPVertModule}{\paperheight}
\TPMargin{5pt}
\begin{textblock}{0.8}(0.1,0.042)
     \noindent
     \footnotesize
     If you cite this paper, please use the original reference:
     Timo H\"ackel, Philipp Meyer, Franz Korf, and Thomas C. Schmidt. Secure Time-Sensitive Software-Defined Networking in Vehicles. \emph{IEEE Transactions on Vehicular Technology}, IEEE, 2022. \url{https://doi.org/10.1109/TVT.2022.3202368}
\end{textblock}

\begin{abstract}
 Current designs of future In-Vehicle Networks (IVN) prepare for switched Ethernet backbones, which can host advanced LAN technologies such as IEEE Time-Sensitive Networking (TSN) and  Software-Defined Networking (SDN).
  In this paper, we present an integrated Time-Sensitive Software-Defined Networking (TSSDN) architecture that simultaneously enables control of  synchronous and asynchronous real-time and best-effort communication for all IVN traffic classes. 
	Despite the central SDN controller, we can validate that control can operate without a delay penalty for TSN traffic, provided protocols are properly mapped.
  We demonstrate how TSSDN  adaptably and reliably enhances network security for in-vehicle communication.
  A systematic investigation of the possible control flow integrations with switched Ether-networks reveals
that these strategies allow for shaping the attack surface of a software-defined IVN.
  We discuss embeddings of  control flow identifiers on different layers, covering the range from a fully exposed mapping to deep encapsulation.
  We experimentally evaluate these strategies in a production vehicle, which we map to a modern Ethernet topology.
  Our findings indicate that visibility of automotive control flows on lower network layers enables isolation and access control throughout the network infrastructure. Such a TSSDN backbone can establish and survey trust zones within the IVN and reduce the attack surface of connected cars in various attack scenarios.
\end{abstract}
\vspace{-4mm}

%
\IEEEpeerreviewmaketitle

\begin{acronym}
	\acro{ACC}[ACC]{Adaptive Cruise Control}
	\acro{ACL}[ACL]{Access Control List}
	\acro{ADS}[ADS]{Anomaly Detection System}
	\acroplural{ADS}[ADSs]{Anomaly Detection Systems}
	\acro{ADAS}[ADAS]{Advanced Driver Assistance Systems}
	\acro{API}[API]{Application Programming Interface}
	\acro{AVB}[AVB]{Audio Video Bridging}
	\acro{ARP}[ARP]{Address Resolution Protocol}
	\acro{BE}[BE]{Best-Effort}
	\acro{CAN}[CAN]{Controller Area Network}
	\acro{CBM}[CBM]{Credit Based Metering}
	\acro{CBS}[CBS]{Credit Based Shaping}
	\acro{CNC}[CNC]{Central Network Controller}
	\acro{CMI}[CMI]{Class Measurement Interval}
	\acro{CoRE}[CoRE]{Communication over Realtime Ethernet}
	\acro{CT}[CT]{Cross Traffic}
	\acro{CM}[CM]{Communication Matrix}
	\acro{DoS}[DoS]{Denial of Service}
	\acro{DDoS}[DDoS]{Distributed Denial of Service}
	\acro{DPI}[DPI]{Deep Packet Inspection}
	\acro{ECU}[ECU]{Electronic Control Unit}
	\acroplural{ECU}[ECUs]{Electronic Control Units}
	\acro{GCL}[GCL]{Gate Control List}
	\acro{HTTP}[HTTP]{Hypertext Transfer Protocol}
	\acro{HMI}[HMI]{Human-Machine Interface}
	\acro{IA}[IA]{Industrial Automation}
	\acro{IDS}[IDS]{Intrusion Detection System}
	\acroplural{IDS}[IDSs]{Intrusion Detection Systems}
	\acro{IEEE}[IEEE]{Institute of Electrical and Electronics Engineers}
	\acro{IoT}[IoT]{Internet of Things}
	\acro{IP}[IP]{Internet Protocol}
	\acro{ICT}[ICT]{Information and Communication Technology}
	\acro{IVNg}[IVN]{In-Vehicle Networking}
	\acro{IVN}[IVN]{In-Vehicle Network}
	\acroplural{IVN}[IVNs]{In-Vehicle Networks}
	\acro{LIN}[LIN]{Local Interconnect Network}
	\acro{MOST}[MOST]{Media Oriented System Transport}
	\acro{NADS}[NADS]{Network Anomaly Detection System}
	\acroplural{NADS}[NADSs]{Network Anomaly Detection Systems}
	\acro{OEM}[OEM]{Original Equipment Manufacturer}
	\acro{P4}[P4]{Programming Protocol-independent Packet Processors}
	\acro{PCP}[PCP]{Priority Code Point}
	\acro{RC}[RC]{Rate-Constrained}
	\acro{REST}[ReST]{Representational State Transfer}
	\acro{RPC}[RPC]{Remote Procedure Call}
	\acro{SDN}[SDN]{Software-Defined Networking}
	\acro{SDN4CoRE}[SDN4CoRE]{Software-Defined Networking for Communication over Real-Time Ethernet}
	\acro{SOA}[SOA]{Service-Oriented Architecture}
	\acro{SOME/IP}[SOME/IP]{Scalable service-Oriented MiddlewarE over IP}
	\acro{SR}[SR]{Stream Reservation}
	\acro{SRP}[SRP]{Stream Reservation Protocol}
	\acro{SW}[SW]{Switch}
	\acroplural{SW}[SW]{Switches}
	\acro{TAS}[TAS]{Time-Aware Shaping}
	\acro{TCP}[TCP]{Transmission Control Protocol}
	\acro{TDMA}[TDMA]{Time Division Multiple Access}
	\acro{TSN}[TSN]{Time-Sensitive Networking}
	\acro{TSSDN}[TSSDN]{Time-Sensitive Software-Defined Networking}
	\acro{TT}[TT]{Time-Triggered}
	\acro{TTE}[TTE]{Time-Triggered Ethernet}
	\acro{UDP}[UDP]{User Datagram Protocol}
	\acro{QoS}[QoS]{Quality-of-Service}
	\acro{V2X}[V2X]{Vehicle-to-X}
	\acro{WS}[WS]{Web Services}
	\acro{ZC}[ZC]{Zone Controller}

\end{acronym}

\section{Introduction}
\label{sec:introduction}
%
%
%
%


\IEEEPARstart{V}{ehicles} continuously implement new features based on sensors and actuators connected with \acp{ECU}.
Traditionally, the \ac{IVN} is organized in functional domains using a combination of bus systems. 
Automotive Ethernet has emerged as the next high-bandwidth communication technology~\cite{mk-ae-15}.
Future \acp{IVN} will migrate to switched Ethernet~\cite{brkw-aeaet-17} as shared backbones for different domains and service requirements.
For time-constraint traffic, the standards of \acf{TSN} (IEEE 802.1Q~\cite{ieee8021q-18}) add real-time capabilities to Ethernet.

\ac{ADAS} and similar features increase cross-domain communication and functions implemented in software following a \ac{SOA}.
These services are highly dynamic and require an adaptable \ac{IVN}.
\ac{SDN}~\cite{mabpp-oeicn-08} has been identified as a powerful building block for \acp{IVN}, as it promises to increase robustness and adaptability~\cite{hmg-rsarn-18, hmks-snsti-19, hhlng-saeea-20}.
In \ac{SDN}, the control plane of the network devices is offloaded to a central controller. 
On the data plane, network devices forward packets based on pipelines controlled by the \ac{SDN} controller.
\ac{TSSDN} was introduced to enable centralized reconfiguration of time-sensitive communication~\cite{ndr-tssdn-16}.
In recent work, we integrated \ac{TSN} with \ac{SDN} to control asynchronous real-time traffic using the OpenFlow protocol without a delay penalty~\cite{hmks-snsti-19}.

Security challenges arise from communication with other vehicles or roadside units (\acs{V2X}), and via Internet uplinks that open a vehicle to remote attackers.
All this, and also the flattening interconnect of domains increase the vulnerability of safety-critical functions and require versatile measures to secure future vehicles~\cite{m-saejr-17}.
Current vehicles are vulnerable to manipulation by third parties, which became apparent through cyber-attacks in the field~\cite{mv-reupv-15}.
A robust \ac{IVN} can limit the attack surface and reduce the impact on communication.

\ac{SDN} can improve network security by separating the control from the data plane~\cite{sxhg-enssd-16}.
The central controller has global network knowledge and devices only forward packets according to predefined flows. 
This bears the potential to detect, prevent, or mitigate cyber-attacks~\cite{yd-sbcjr-21}.
In previous work, we identified the \ac{SDN} matching pipeline as a powerful tool to accurately identify, separate, and protect native \ac{IVN} communication~\cite{hsmks-sicfs-20}. 

This paper develops \ac{TSSDN} further by integrating transactional updates for synchronous real-time traffic and thereby completes the set of fully programmable options for controlling all classes of in-vehicle communication via a central \ac{SDN} controller.
Applying control programming to critical real-time traffic significantly extends network security mechanisms for in-vehicle communication.
Our main contributions read:
\begin{enumerate}
    \item We integrate \ac{TSN} with \ac{SDN} in a switching architecture that enables central monitoring and control for all classes of \ac{IVN} communication without delaying real-time traffic.
    \item We present and evaluate an approach to secure in-vehicle TSSDN by introducing reliable static configuration and secure adaptive communication.
    \item We evaluate different approaches to embed in-vehicle control flows in SDN and quantify the precision with which the network can identify and isolate them.
    \item We show that TSSDN can isolate in-vehicle control flows in a shared environment,  prevent unwanted traffic, and significantly reduce the attack surface in a prototype built from a production vehicle. 
\end{enumerate}

The remainder of this work is organized as follows.
Section~\ref{sec:background_related_work} reviews the \ac{IVN}, \ac{TSN}, and \ac{SDN} together with related work on network security in cars. 
Section~\ref{sec:tssdn} introduces the concept of \ac{TSSDN}.
\ac{SDN}-enabled security measures for \acp{IVN} are presented in Section~\ref{sec:security_concepts}.
Section~\ref{sec:eval_validation} validates the \ac{TSSDN} architecture in simulations.
Section~\ref{sec:eval_security} analyses the security impact of our access-control in a prototype car and investigates benefits and limits in attack scenarios.
Finally, Section~\ref{sec:conclusion_outlook} concludes our paper with an outlook on future work.

\section{Background and Related Work}%
\label{sec:background_related_work}
\subsection{Evolution of In-Vehicle Networks}
\label{subsec:bgivn}
Modern \acp{IVN} connect sensors, actuators, and \acp{ECU}.  
Some in-vehicle communication is static and well-defined, such as motor control, while other is dynamic and not always active, such as \ac{ADAS}. 
All communication relations between the installed \acp{ECU} are specified in a communication matrix.
Each control flow in the matrix has exactly one sender, a list of receivers, and a unique identifier across the vehicle.

\subsubsection{Traffic types}
In-vehicle communications include periodic control messages, sporadic events, and media traffic. 
Besides the vehicle-specific traffic, there is also network control, which is required for reconfiguration and service discovery, and \acf{BE} communication.

The different traffic classes have varying \ac{QoS} requirements. 
Control communication often uses small frames in control loops with a fast period and may be very susceptible to jitter which requires fixed latencies around \SI{100}{\micro\second}. 
Media traffic (of e.g., LIDAR systems) may require several \si{\giga\bit\per\second} with guaranteed latencies around \SI{10}{\milli\second}.
Safety-critical messages may have no tolerance for packet loss and require redundancy and hard deadlines.
For some applications, retransmission may compensate for packet loss, while with cyclic control communication, retransmission is irrelevant. 

\subsubsection{Network topology}
\begin{figure}
    \vspace{-7pt}
    \centering
    \subfloat[Central Gateway]{\includegraphics[width=.3\linewidth, trim=4.3cm 4.3cm 4.3cm 4.3cm, clip=true]{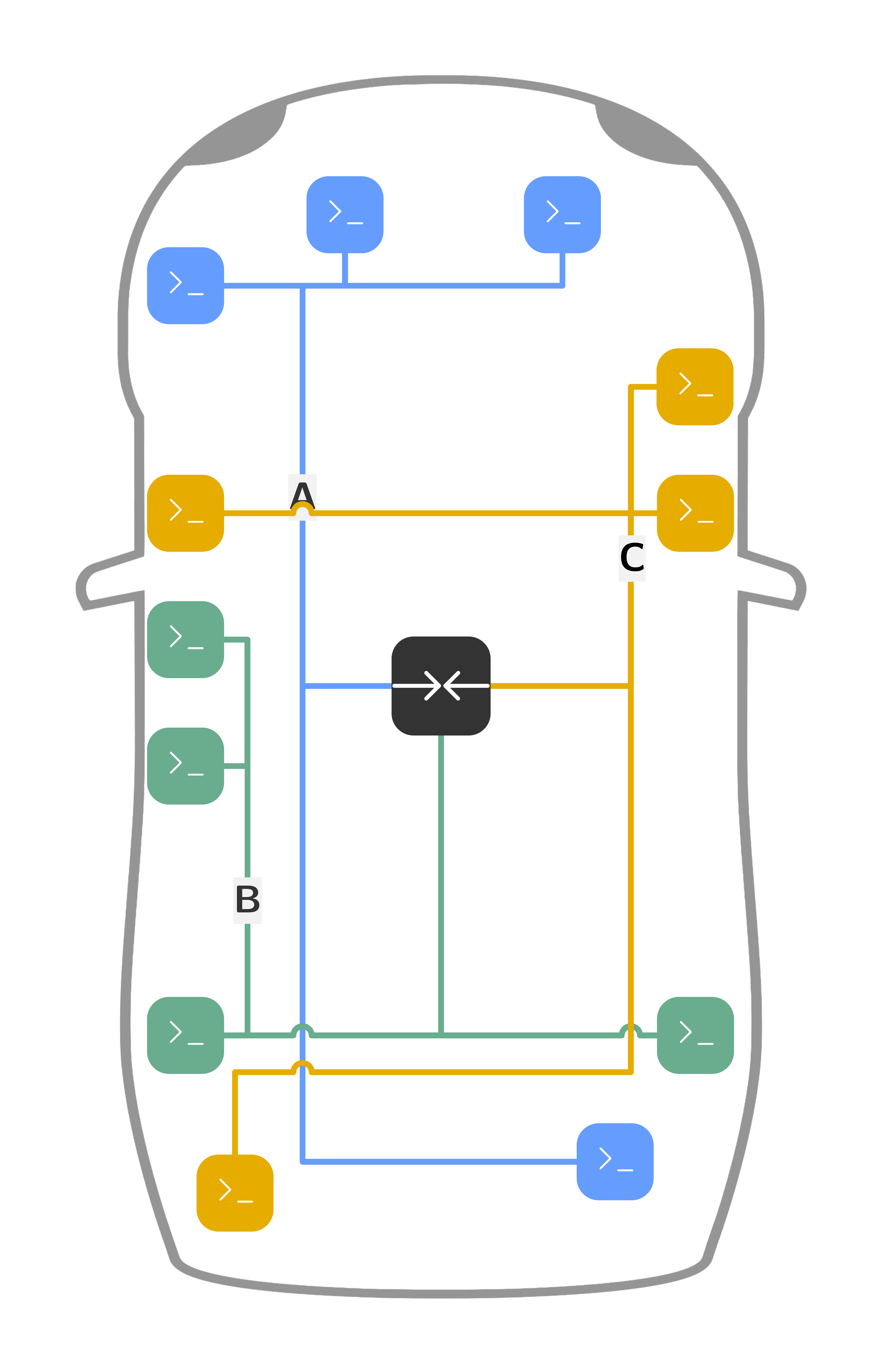}%
    \label{fig:topo_central}}
    \hfill
    \vrule height 125pt \relax
    \hfill
    \subfloat[Domain Controller]{\includegraphics[width=.3\linewidth, trim=4.3cm 4.3cm 4.3cm 4.3cm, clip=true]{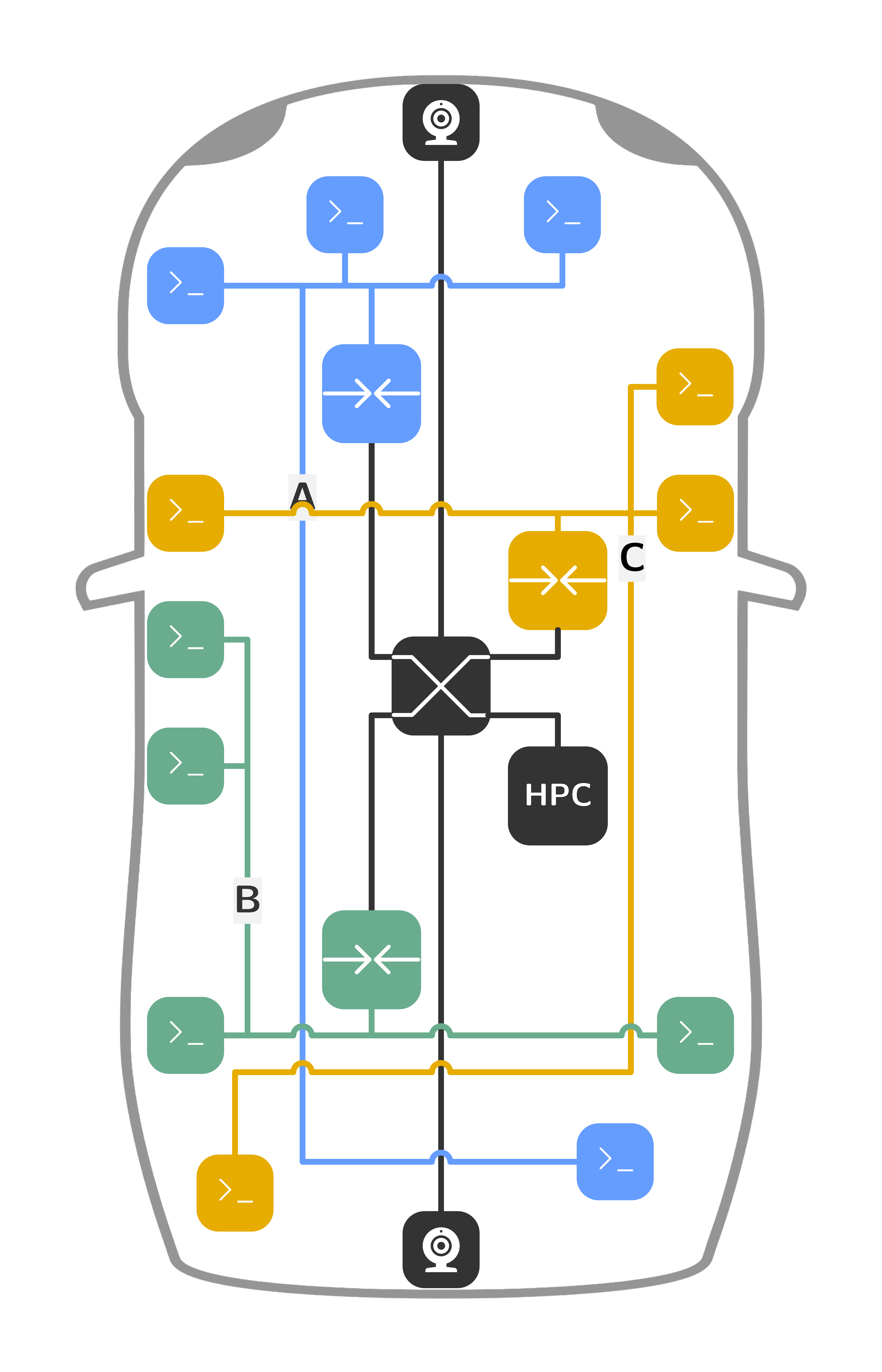}%
    \label{fig:topo_domain}}
    \hfill
    \vrule height 125pt \relax
    \hfill
    \subfloat[Zone Controller]{\includegraphics[width=.3\linewidth, trim=4.3cm 4.3cm 4.3cm 4.3cm, clip=true]{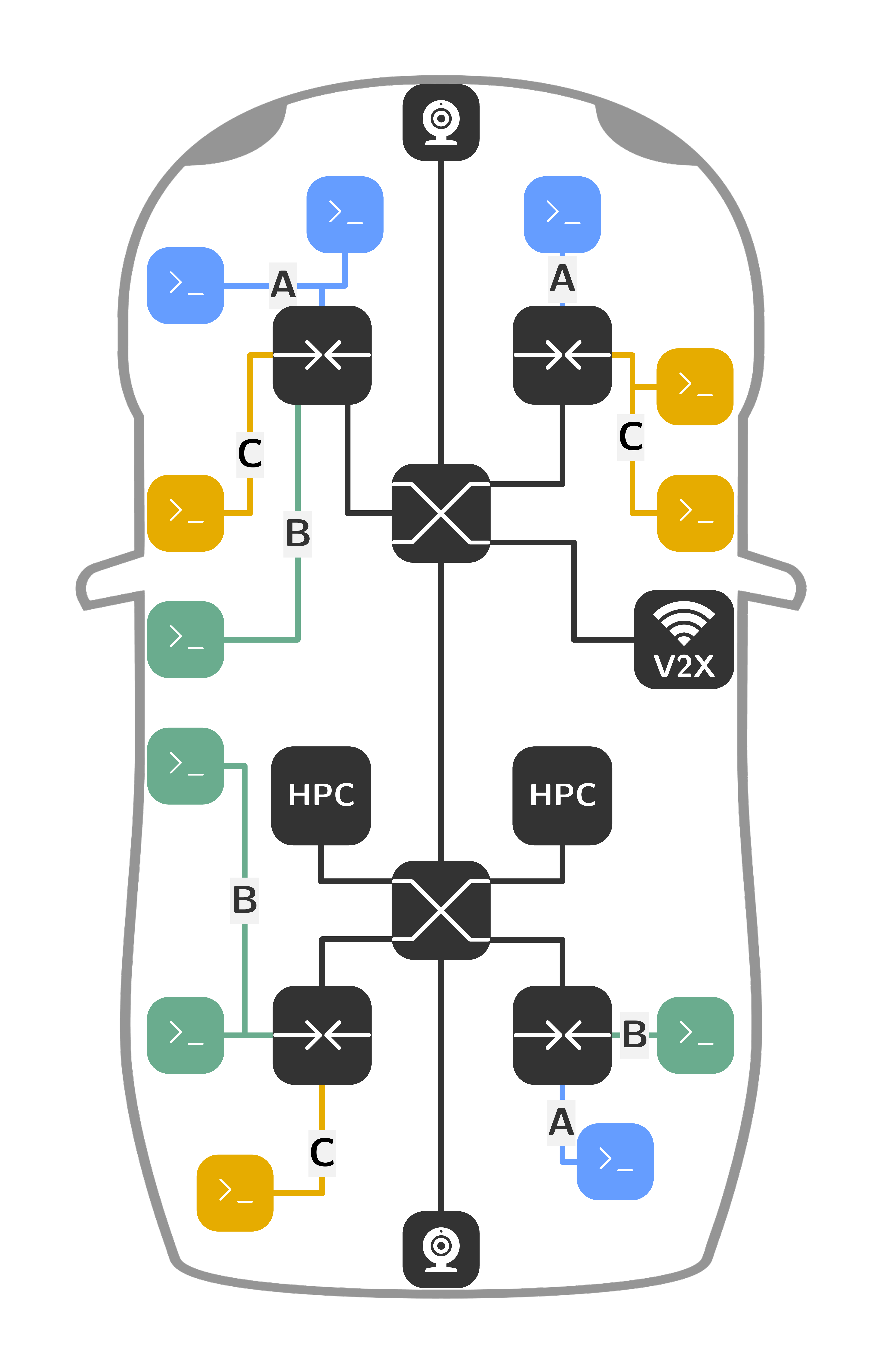}%
    \label{fig:topo_zonal}}

    \includegraphics[width=0.85\linewidth, trim=0.6cm 0.8cm 0.7cm 0.7cm, clip=true]{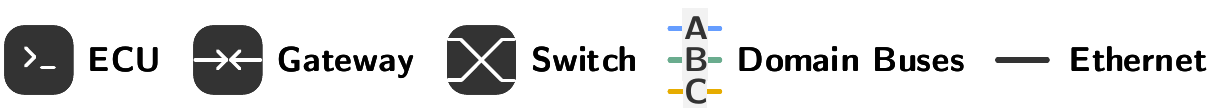}
    \caption{
    Evolution of the IVN from a central gateway topology that enables cross-domain communication, to a domain controller topology that connects network domains via an Ethernet backbone, to a zone topology that splits the domain buses and connects all ECUs to a zone controller in their vicinity.
    }
    \label{fig:topologies}
    \vspace{-2mm}
\end{figure}

IVN topologies have evolved to adapt to the increasing communication demands of automotive applications~\cite{dzjal-adijr-20, brkw-aeaet-17, mk-ae-15, m-saejr-17}.
Initially (Figure~\ref{fig:topo_central}), \acp{ECU} were grouped into functional domains such as chassis control, powertrain, comfort, and infotainment using heterogeneous bus systems such as \ac{CAN}, \ac{MOST} and \ac{LIN}, which physically extended over large areas of the vehicle.
For cross-domain functions, a central gateway was installed to transfer messages from one domain to another. 

Today, cross-domain communication is increasingly required to enable features such as \ac{ADAS}, and new sensors such as high-resolution cameras need high bandwidths for communication.
Figure~\ref{fig:topo_domain} shows a domain controller topology which uses a switched Automotive Ethernet~\cite{mk-ae-15} backbone that enables fast cross-domain communication.
Domain gateways integrate legacy devices and forward messages between the domain buses and the Ethernet backbone.
Central compute units were introduced as `High-Performance Computers (HPCs)'. 
They offer much higher processing power than traditional \acp{ECU} and bundle virtual functions of computationally intensive tasks.

Future vehicles will communicate with other vehicles, roadside units (\acs{V2X}) and the Internet.
More and more functions will be implemented in software and the number of OEM model variants will increase.
This requires higher flexibility of software components and increases the dynamics of the network.
The zone model~(\ref{fig:topo_zonal}) connects all \acp{ECU} with a zone controller in physical proximity (e.g., front left), so that wiring can be reduced.
\mbox{AUTOSAR} paves the way for the transition to \ac{SOA} with \acf{SOME/IP}~\cite{autosar-someip-17}.
Containerized services are under discussion for larger \acp{ECU}~\cite{kakww-dsosa-19}.
This poses challenges for the network as safety-critical traffic shares a wire with other traffic. 

\vspace{-10pt}
\subsection{\acl*{TSN} in Cars}
\label{subsec:bgtsn}
In-vehicle communication requires robust \ac{QoS} for simultaneous real-time and \ac{BE} traffic.
There are several proposals to add real-time capabilities to Ethernet, of which the standards for \acf{TSN} ({IEEE 802.1Q}~\cite{ieee8021q-18}) are the most promising candidate for deployment in vehicles~\cite{ls-pitjr-19}.

\subsubsection{Real-time control}
On the input and output ports of \ac{TSN} devices, gates can be opened to let frames through, or closed to block them.
A periodic \acf{GCL} schedules at what time a gate opens and closes. 
Meters at the ingress and shapers at the egress complement the \ac{GCL} functionality.

Ingress control (IEEE 802.1Qci) filters incoming frames per stream and already discards packets that missed their time slot. 
For frames that pass the ingress control, the switching fabric decides to which output ports they are forwarded.
Egress control (IEEE 802.1Qbv) assigns frames to one of 8 priority queues, each of which has a gate and may have a selection algorithm, e.g., for bandwidth control.
The first frame from the queue with the highest priority whose selection algorithm and gate allow the frame to pass is selected for transmission.

\subsubsection{Communication classes}
Real-time communication can be synchronous (\ac{TDMA}) or asynchronous as already defined in the \ac{TSN} predecessor \ac{AVB}, which we compared in former work~\cite{slksh-tiice-12}.
Asynchronous real-time communication requires bandwidth reservation of a predefined data rate per flow.
The bandwidth usage is controlled by the selection algorithm of each queue at the egress, e.g., by \ac{CBS}.
Synchronous real-time communication can be implemented with a network-wide \ac{TDMA} schedule for the \acp{GCL}.
This can minimize latency and jitter of real-time traffic, but requires high-precision time synchronization (802.1AS-2020).

\subsubsection{Schedule configuration}
Scheduling in \ac{TSN} has been explored in the past~\cite{gzrp-aarjr-18} and applied to \acp{IVN}~\cite{salc-mbjjr-20}.
In a scheduled network, all endpoints and switches implement \ac{TDMA}.
Schedules can include any selection of priorities, allowing concurrent synchronous and asynchronous communication.

The calculation of a \ac{TDMA} schedule is complex and computationally intensive, therefore schedules are usually computed offline. 
Still, a schedule can be updated during runtime~\cite{salc-mbjjr-20}.
The NETCONF protocol (RFC~6241~\cite{RFC-6241}) can be used with YANG data models (RFC~6020) for reconfiguration of \ac{TSN} modules (e.g., 802.1Qcp and P802.1Qcw).

There are still open challenges, such as simplifying reconfiguration, and improving security to detect threats and initiate countermeasures~\cite{ls-pitjr-19}.
In this paper, we introduce dynamic traffic control for \ac{TSN} by integrating it with \ac{SDN} and propose security mechanisms to protect in-vehicle communication from interference with unknown traffic.

\vspace{-2mm}
\subsection{\acl*{SDN} in Cars}
\label{subsec:bgsdn}
\ac{SDN}~\cite{mabpp-oeicn-08} separates the control logic (control plane) from the underlying switches that forward the traffic (data plane)~\cite{krvra-sdncs-15}.
Network devices become simple forwarders that are programmed by a central \ac{SDN} controller with global network knowledge using open standards such as the OpenFlow protocol~\cite{onfts025-15}.
Controller applications implement the behavior of the network, e.g., routing protocols.
OpenFlow switches forward incoming packets based on a programmable flow table.
A flow entry matches a subset of Layer~2 to Layer~4 header fields and contains actions, such as discard, forward, or modify. 

\subsubsection{Real-time capability}
\ac{SDN} is considered generally suitable for real-time environments in terms of network configuration latency~\cite{te-asdts-16}, in particular when all flows are predefined in the switches.
Still, \ac{SDN} needs to be extended to control bandwidth reservation and scheduling for real-time flows.

Nayak et al.~\cite{ndr-tssdn-16, ndr-ifsjr-18} first mention \acf{TSSDN} with dynamic scheduling and routing techniques during runtime to improve robustness in \ac{TSN}, but only the hosts are scheduled and not yet the switches.
Thereby cross traffic can still impact the performance of real-time traffic classes, as we investigated in former work~\cite{slksh-bhcan-15}.
We argue that a vital requirement for \ac{TSSDN} is to adapt  configurations of network devices to changing real-time traffic.
Earlier~\cite{hmks-snsti-19}, we integrated \ac{SDN} and \ac{TSN} for stream reservation with OpenFlow without delay penalty for time-sensitive in-vehicle communication. 
Gerhard et al.~\cite{gkbh-sdfjr-19} implemented a similar concept in a hardware environment of industrial plants.
Nam et al.~\cite{nkm-ssrjr-21} optimized \ac{TSN} stream reservation with \ac{SDN} and could reduce the communication overhead for stream reservation.
Corresponding work has focused on either \ac{TSN} scheduling techniques (e.g.,~\cite{salc-mbjjr-20}) or dynamic reservation with the SRP. 
In this paper, we extend our approach to \ac{TSSDN} by closing the gap left when updating the schedule in network devices for synchronous traffic. This work completes the design of a fully programmable \ac{TSSDN} for \acp{IVN}.

Transactions are commonly used to coordinate critical changes in distributed systems. 
The ACID properties (atomicity, consistency, isolation, and durability) guarantee consistency of transactions despite possible errors.
ACID transactions can be used in \ac{SDN} environments to maintain valid state of the network~\cite{czzxs-tbfjr-18, cdhc-tnujr-18}.
The impact of transactional network updates on real-time traffic has not been investigated yet.
We compare two methods for transactional network updates in real-time systems, which both maintain a consistent state across distributed network devices.

\subsubsection{Automotive use-cases}
\ac{SDN} promises to reduce complexity and increase the adaptability of networks~\cite{krvra-sdncs-15}.
Halba et al.~\cite{hmg-rsarn-18} showed how \ac{SDN} can improve the safety and robustness of \acp{IVN} through dynamic rerouting.
Haeberle et al.~\cite{hhlng-saeea-20} presented an \ac{IVN} concept that reduces the complexity of the vehicular E/E architecture based on \ac{SDN}.
In previous work~\cite{rhmks-rapesc-20}, we evaluated the performance of existing \ac{SDN} controller implementations with respect to \ac{IVN} requirements and found that all implementations lack important safety requirements, such as guaranteed response times, but could confirm that \ac{SDN} controllers can be used in vehicles with the right modifications.
To the best of our knowledge, no related work has analyzed the use of \ac{SDN} to secure the \ac{IVN}.

In this paper, we investigate how \ac{SDN} flow control can improve the security of the \ac{IVN} by precisely separating in-vehicle communication. 
Combining \ac{TSN} and \ac{SDN} can ensure that real-time requirements are met.

\vspace{-8pt}
\subsection{\acl*{IVN} Security}
\label{subsec:bgsecurity}
Current vehicles are vulnerable to manipulation by third parties, which has been demonstrated in the field~\cite{mv-reupv-15}. 
Checkoway et al.~\cite{cmkas-ceaas-11} provide a fundamental analysis of the automotive attack surface and systematically show how a variety of interfaces can be used to gain malicious access to in-car devices.
Manipulation of the \ac{IVN} and its \acp{ECU} can compromise the safety of the vehicle, putting passengers at risk.

\subsubsection{Security assessment}
Assessing security mechanisms is difficult because the risk of unknown vulnerabilities is hard to predict~\cite{wjscn-kzdsm-14}.
A common method for evaluating \ac{IVN} security is to analyze dark-side scenarios~\cite{rwwir-srads-09}, which are based on known vulnerabilities and attack targets.
These can be used to assess the probability, severity, and controllability of attacks~\cite{mbzls-ssajr-18}.
Based on an attack tree created from real incidents, Longari et al.~\cite{lccz-sdfra-19} evaluated how \ac{CAN} network topologies can be hardened by introducing additional gateways.
In this work, we systematically analyze the separation of in-vehicle control flows in Ether-networks and use attack scenarios to show the benefits and limitations of our network security mechanisms.

\begin{figure*}[ht]
    \centering
    \includegraphics[width=1.0\linewidth, trim= 0.7cm 0.8cm 0.8cm 0.8cm, clip=true]{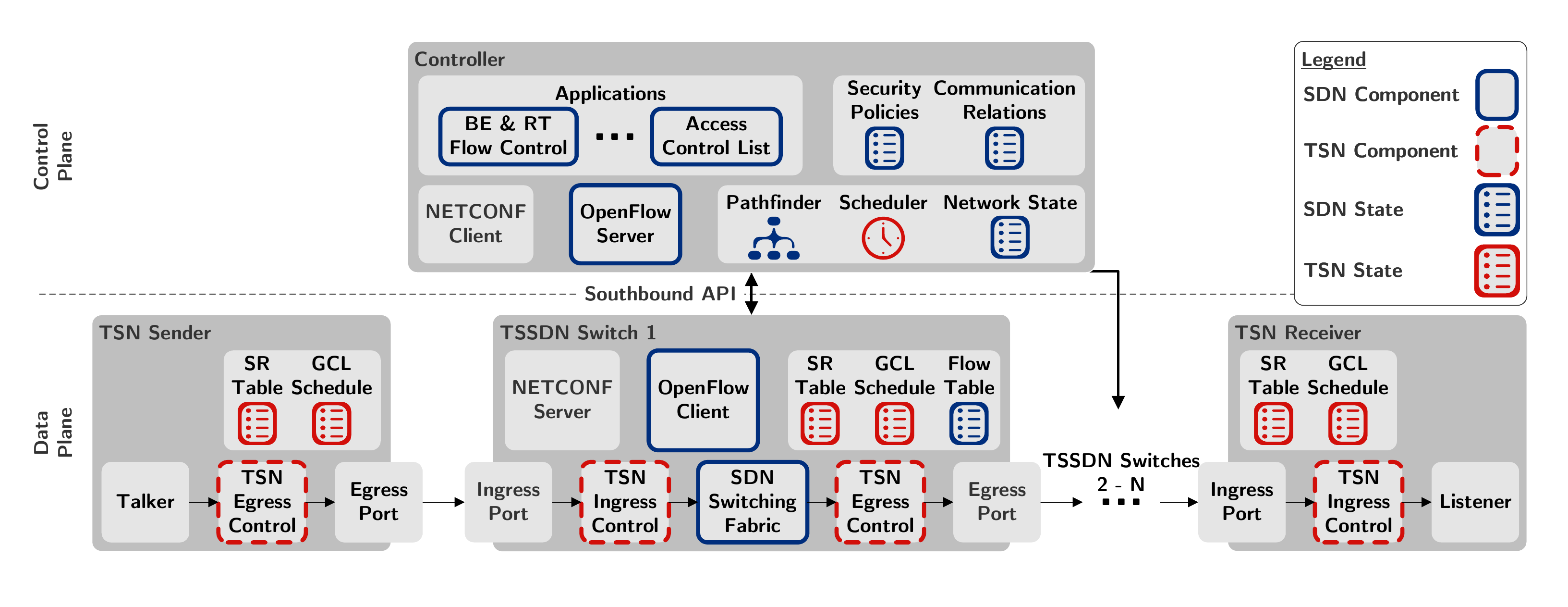}
    \caption{
        Integrated network architecture of TSN and SDN. 
        At the \textbf{data plane}, packets flow from a sending to a receiving TSN endpoint through switches that combine TSN ingress and egress control with SDN forwarding. 
        At the \textbf{control plane}, TSN functions such as scheduling are integrated with common SDN controller functions such as path finding using a global network view. 
        Network applications control real-time and best-effort flows simultaneously and enforce security policies.
    }
    \label{fig:tssdn_architecture}
    \vspace{-8pt}
\end{figure*}

\subsubsection{Taxonomy of attacks and defenses}
Attacks on \acp{IVN} include \ac{DoS}, replay, spoofing, malware, and falsified-information attacks~\cite{tw-avsjr-16, dzjal-adijr-20}.
They can be grouped in \textit{alter} attacks that aim to modify data, \textit{listen} attacks that aim to monitor data, \textit{disable} attacks that aim to deny services, and \textit{forge} attacks that aim to insert incorrect data~\cite{mbzls-ssajr-18}.

Defensive measures can be divided into attack prevention, detection, and mitigation~\cite{tw-avsjr-16}.
The main security goals for \acp{IVN} are 
(i) \textit{availability}, which ensures that resources and services are accessible; 
(ii) \textit{integrity}, which ensures the accuracy and completeness of data;
and (iii) \textit{authenticity}, which aims at the verifiability of data sources and sinks~\cite{mbzls-ssajr-18, dzjal-adijr-20}. 

\subsubsection{Security mechanisms in IVNs}
Firewalls and access control mechanisms can prevent attacks with stateful inspection, rate limiting, and filtering~\cite{psz-hcdjr-17}.
Gateways can filter messages between different sections of the \ac{IVN}~\cite{rdgks-iabac-19} and analyze event chains based on behavior specifications~\cite{so-sagjr-14}.
In this work, we focus on network defense mechanisms that rely on precise \ac{SDN}-based flow separation in the forwarding pipeline without additional processing resources.

Anomaly detection can identify ongoing attacks in the network~\cite{rmwh-sadjr-18}, e.g., using machine learning algorithms to detect anomalies on the \ac{CAN} bus~\cite{ymhs-tbijr-19}.
In previous work~\cite{mhks-nadci-20}, we presented an anomaly detection system using the \ac{TSN} ingress control that can operate with zero false-positives based on a precise traffic specification.
The reported security incidents from different detection mechanisms can be combined~\cite{wmslk-sdvtm-17} and analyzed for entire vehicle fleets in the cloud to detect correlations, e.g., in terms of affected devices~\cite{lss-eacdc-19}. 

Integrity and authenticity can be ensured with authentication and encryption methods suitable for \acp{IVN}~\cite{m-saejr-17, hl-rscan-18}, across all communication systems, such as CAN~\cite{fizj-spsjr-17}, FlexRay~\cite{paak-sfbjr-20}, and Ethernet, e.g., MACSec (IEEE 802.1AE~\cite{ieee8021ae-18}) on the MAC layer.
Identity and credential management enables the verification of entities such as \acp{ECU}~\cite{kjp-sricm-17}.
Encryption, however, cannot protect  data flows from all network attacks.
In-vehicle \acp{ECU} have limited computing power which limits the resources for security features.
In addition, established and verified \acp{ECU} are used over many vehicle generations.
Our network-centric approach helps to secure \acp{ECU} with limited computing power and allows secure reuse of legacy devices while remaining compatible to application layer encryption.

\subsubsection{Software-Defined Networking}
Flexible security solutions will be beneficial to cope with the growing dynamics of the \ac{IVN}~\cite{rgks-oasjr-20}. 
Software-defined security elements can be easily adapted to the situation at hand~\cite{dzjal-adijr-20}.
\ac{SDN} can improve network security by detecting and preventing attacks~\cite{yd-sbcjr-21}.
In previous work~\cite{hsmks-sicfs-20}, we showed how the precise flow control of \ac{SDN} can be used to protect \ac{IVN} control communication that is precisely specified in the communication matrix.

In this work, we use \ac{SDN} mechanisms that protect \ac{IVN} communication from malicious access.
Enabling detection of unknown flows and countermeasures through reconfiguration.

\section{\acl{TSSDN}} 
\label{sec:tssdn}
\acf{TSSDN} integrates the \ac{TSN} and \ac{SDN} concepts on the control and data plane as shown in Figure~\ref{fig:tssdn_architecture}. 
On the data plane, forwarding follows \ac{TSN} ingress and egress control together with the \ac{SDN} flow switching rules. 
On the control plane, real-time traffic control commands for switches are perceived and processed by the \ac{SDN} controller. 
An open southbound API enables vendor-independent programming of \ac{TSN} and \ac{SDN} components. 

\vspace{-8pt}
\subsection{Data Plane Architecture}
The \ac{TSSDN} data plane connects the \ac{TSN} endpoints via switches that integrate the \ac{SDN} forwarding pipeline with the \ac{TSN} real-time control. 
Each switch contains a flow table, an \ac{SR} table, and a \acf{GCL}. 

Frames arrive at the \ac{TSN}-controlled ingress, which applies filters and time checks. 
After a frame passed the ingress control, the \ac{SDN} switching fabric performs a lookup in its flow table for forwarding. 
If a matching entry is found, the predefined actions are performed and the packet is forwarded to its specified ports. 
The \ac{TSN} egress control of the port then shapes the outbound traffic.
If no matching rule exists, the packet is discarded by default.
Most controllers, however, install a rule to receive these packets. 

\vspace{-8pt}
\subsection{Control Plane Operations}
The \ac{SDN} paradigm mandates that control plane functions are offloaded from the switches to a central controller. 
The controller combines common \ac{SDN} tasks such as address learning and routing with 
\ac{TSN} functions such as stream reservation and scheduling. 
Network designers implement applications to control the behavior of the network through the controller using OpenFlow or NETCONF.
A network application can react to messages and push updates to the data plane, thereby leveraging the abstract network view, and knowledge about the \ac{IVN} as described by the communication matrix and security policies. 
TSN traffic differentiation will operate as follows. 

\subsubsection{Best-effort flow control}
\ac{BE} flow control remains unaltered to traditional \ac{SDN}. 
OpenFlow-enabled switches forward packets of unknown flows to the controller. 
Network applications decide whether to discard the packet, reply directly, or forward it.
For the latter, the application determines a route and installs flow rules on the data plane.
Thereafter, network devices can forward packets of this flow independently.

\subsubsection{Asynchronous real-time flow control}
For asynchronous real-time flows, senders and receivers announce their resource requirements across the network using the \acf{SRP}. 
Originally, \ac{TSN} uses a fully distributed control plane of the switches. 
Talkers announce streams by a \textit{Talker Advertise}, which contains information about the stream and its bandwidth demands. 
Switches update their SR table and re-broadcast the announcement. 
Hosts willing to subscribe to a stream send a \textit{Listener Ready} to the talker, and all devices along the path reserve bandwidth for the stream if available.

The conceptual architecture of the centralized stream reservation model (802.1Qcc) harmonizes  well with the \ac{SDN} paradigm.
A \acf{CNC} signals the stream reservation while the communication mechanism between the controller and the network devices is not specified.
We map this centralized model onto the OpenFlow protocol, detailed in~\cite{hmks-snsti-19}.
Again, talkers and listeners announce streams using the \ac{SRP}.
Network devices forward all \ac{SRP} packets to the \ac{SDN} controller.
A network application checks whether the available bandwidth suffices and creates a flow entry that matches the stream. 
Flow entries are updated as listeners leave or join.
Our controller implementation uses an OpenFlow experimenter extension to reserve bandwidth in the \ac{SR} table on the switches. 
With this, forwarding devices can identify the stream, forward it correctly, and control the bandwidth on the egress ports, e.g., with \acf{CBS}. 

\subsubsection{Synchronous real-time flow control}
Synchronous real-time flows are coordinated between different transmitters across multiple links. 
Senders periodically transmit frames of a known maximum size in their time slot. 
A periodic \acf{GCL} schedule opens and closes specific priority gates of each output port. 
Time slots are shifted for devices along the path according to the packet transmission delay, which enables minimal end-to-end latency and jitter, but requires high-precision time synchronization (802.1AS-2020). 

The complex calculation of \ac{TDMA} schedules is commonly performed offline.
Such a static schedule is not efficient for bandwidth usage, since bandwidth remains reserved even if synchronous services are not running. 
In addition, communication changes are not supported but may occur after updating or transitioning applications between devices.

In TSSDN, the controller can dynamically (re)calculate the \ac{GCL} schedule and paths for all flows when synchronous traffic changes. 
Table~\ref{tab:operations} lists the four basic operations of changing the network configuration along with its mandatory order of execution, which we discuss in detail below.
Time slots can be added or removed, and moved forward or backward within the period so that traffic is transmitted earlier or later.
More complex operations can combine these operations, such as rerouting a flow (add and remove) or shifting time slots to make room for a new flow (shift and add).

\begin{table}
    \centering
    \setlength{\tabcolsep}{5pt}
    \caption{
        Basic operations for configuring scheduled traffic and their execution order on network devices along the path.
    }
    \label{tab:operations}
    \begin{tabularx}{\linewidth}{lX}
        \toprule
        \textbf{Schedule update basic operation} &\textbf{Order of execution}
        \\ \midrule
        Add flow and new time slot & 
        From destination to source
        \\
        Remove flow and existing time slot & 
        From source to destination
        \\
        Shift time slot to earlier point in the period & 
        From source to destination
        \\
        Shift time slot to later point in the period  & 
        From destination to source
        \\
        \bottomrule
    \end{tabularx}
    \vspace{-10pt}
\end{table}

\begin{figure}
    \centering
    \includegraphics[width=0.95\linewidth, trim=0.65cm 5.6cm 0.7cm 0.45cm, clip=true]{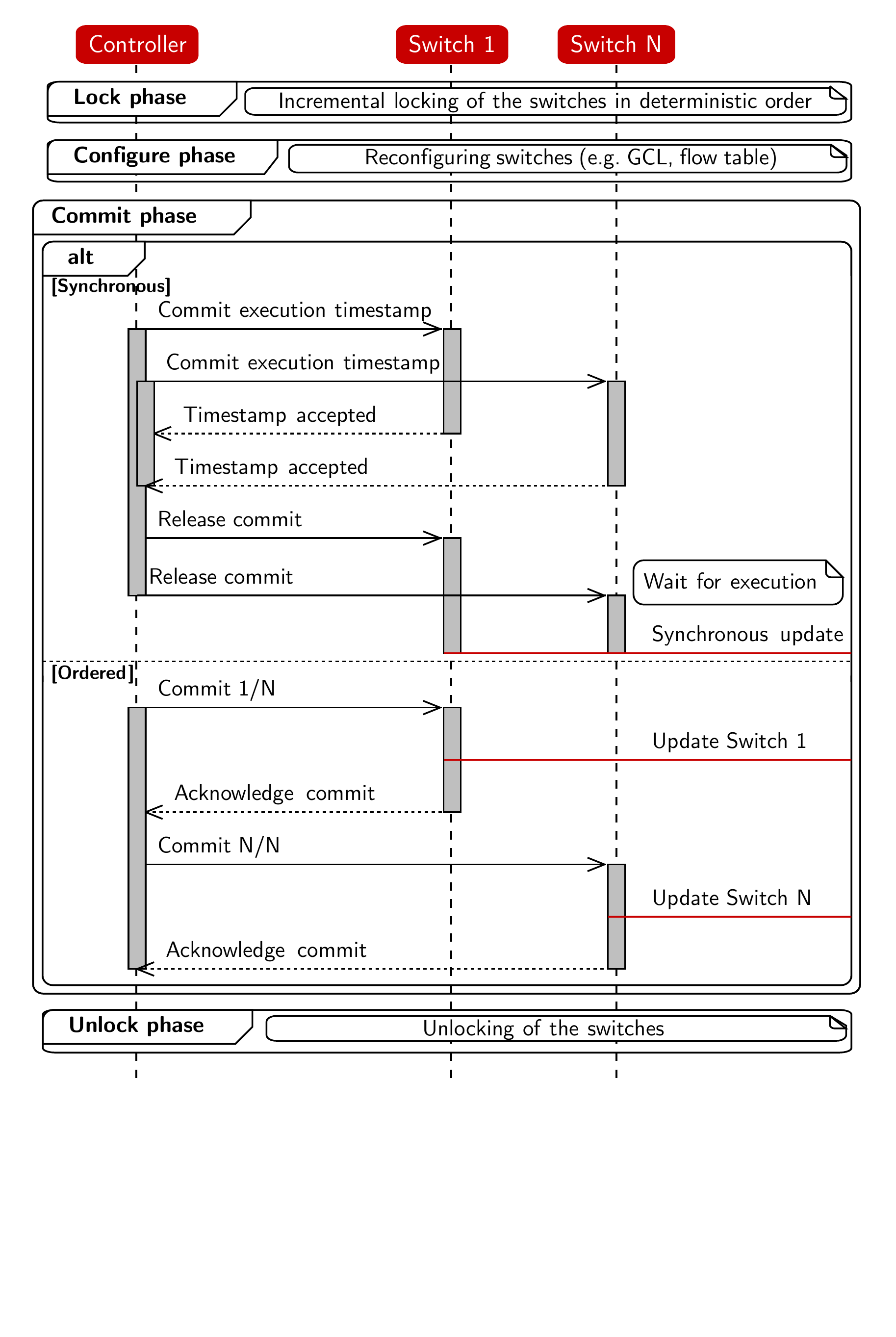}
    \vspace{-6pt}
    \caption{
        Synchronous and ordered network-wide transactional update sequence.
    }
    \label{fig:updatesequence}
    \vspace{-12pt}
\end{figure}

\begin{figure*}
    \centering
    \subfloat[Hidden embedding inside a SOME/IP application layer tunnel --- the network cannot match the hidden control flow identifier.]{\includegraphics[width=1.0\linewidth, 
    trim=0.62cm 4.5cm 0.62cm 0.62cm, clip=true
    ]{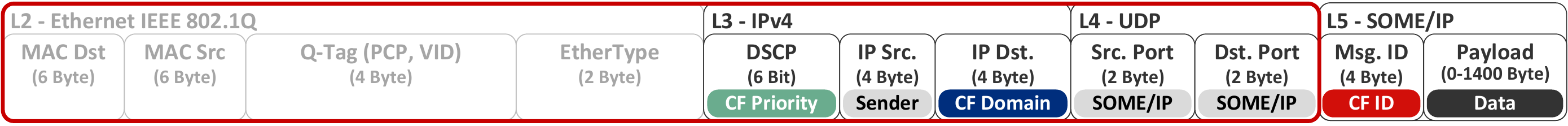}%
    \label{fig:embeddingsomeip}}  

    \vspace{-8pt}

    \subfloat[Exposed embedding in the Ethernet header --- the network can match the full control flow context.]{\includegraphics[width=1.0\linewidth, 
    trim=0.62cm 1.2cm 0.62cm 3.95cm, clip=true
    ] {fig4_embeddingstrategies.pdf}%
    \label{fig:embeddingl2}}  

    \vspace{-4pt}

    \caption{
        Different strategies for embedding Control Flow (CF) context including the identifier (ID), domain, and priority on different layers. 
        The red outline indicates the matchable header fields in the network.
    }
    \label{fig:embeddings}
    \vspace{-8pt}
\end{figure*}

\vspace{-8pt}
\subsection{Transactional Updates for Real-Time Communication}
The \ac{GCL} is scheduled per priority queue, not per flow. 
Adding flows to \ac{TDMA}-scheduled priorities without updating \ac{GCL} 
can lead to overflowing queues and missed deadlines for critical traffic.
Thus, the dynamic nature of \ac{SDN} flow control is inappropriate to control synchronous real-time communication. 
In contrast, the NETCONF protocol is particularly suitable as it supports transaction-oriented configurations.
Device configurations can be locked to guarantee isolation. 
A candidate configuration holds a copy of the device configuration and can be modified and validated before applying the changes. 
If an error occurs, the entire transaction is rolled back, maintaining the previously valid device state; a commit applies the changes to the running configuration otherwise.  

NETCONF transactions, however, only account for operations on a single device. 
Transactions that span multiple devices are required to avoid queue overflow and packet loss during schedule reconfiguration. 
For example, if a time slot is moved to an earlier point in the schedule while a packet is between two network devices, the packet will miss its time slot on the next device. 
The packet is then sent in the next cycle, causing a delay for all upcoming traffic of that priority.

The \ac{TSSDN} controller manages network-wide transactions in four phases (see Figure~\ref{fig:updatesequence}): \textit{lock}, \textit{configure}, \textit{commit}, and \textit{unlock}. 
All devices are locked in deterministic order (e.g., by MAC address) to prevent deadlocks.
Candidate configurations are configured and validated on all devices. 
The commit is then orchestrated and all devices are unlocked afterwards. 
We identified two ways to coordinate a commit across devices consistently without packet loss or delays during configuration.

For \textbf{synchronous updates}, the controller calculates the activation time of the changes and performs a two-stage commit, first ensuring that each switch is ready and will accept the timestamp, then releasing the commit for execution on all devices at the arranged time. 
The timestamp can be a period number, or an absolute, TSN synchronized time. 
It should be chosen so that no scheduled traffic is forwarded in the network devices during the commit. 
This way, synchronous updates can support all combinations of basic operations in one transaction.

For \textbf{ordered updates}, the order of commit execution is determined depending on the reconfiguration, and the controller waits for confirmation of commit execution on one device to commit the next devices. 
Table~\ref{tab:operations} shows the execution order for basic operations. 
Time slots are added from the destination of the new flow towards the source to ensure that no packet enters the network before its time slot has been added on all devices. 
Removing a time slot starts at the source and ends at the destination to ensure that no packets remain on the network. 
To shift a slot ahead in time, the changes proceed from the source to the destination to ensure that packets are early for their time slots rather than too late. 
Shifting to a later point proceeds in reverse order. 
This tells us that the update order is application-specific. 
Updates that combine multiple operations with different commit orders must be split into multiple transactions, which we analyze in Section~\ref{sec:eval_validation}.

Ordered updates rely on standard NETCONF commit management, but synchronous updates also require timing the commit execution. 
In addition, we assume that commit execution on forwarding devices is executed atomically and always succeeds.
Although both approaches guarantee interference-free communication, there are still open problems that remain for future work: 
The start time of a transaction must be set so that all changes proceed before the traffic starts -- we set the time accordingly in our simulation. 
A protocol is missing for senders of synchronous traffic 
to inform the controller about characteristics of their traffic, e.g., packet size, frequency, start and stop times -- we assume that the controller knows them.

\section{Securing Time-Sensitive Software-Defined In-Vehicle Networks}
\label{sec:security_concepts}
For a specific car, critical in-vehicle communication is deterministically defined, which enables the controller to steer traffic precisely based on its \ac{IVN} knowledge including \acp{ECU}, control flows, and timing information from the communication matrix.
\ac{SDN} switches identify flows in the OpenFlow pipeline by matching packet header fields from Layer~2 to Layer~4.

\vspace{-4pt}
\subsection{Embedding Strategies for Control Flow Information} 
\label{sec:layer_embeddings}
A \textbf{Control Flow (CF)} is a sequence of related messages with the same unique identifier in the vehicle -- called a CF ID. 
This could be a CAN message, for example.
A CF is sent from a single origin and follows a point-to-multipoint relationship to reach one or more receivers.
The priority of a CF is determined by its deadline and criticality, e.g., safety critical messages have the highest priority. 

Gateways forward data between bus systems and the Ethernet backbone.
Therefore, the CF is repackaged, e.g., from a CAN frame to an Ethernet frame. 
Future Ethernet communication in vehicles can exploit the entire stack to embed control flows. 
Figure~\ref{fig:embeddings} presents two examples of CF embedding on different layers. 
Each CF has an identifier (ID), a sender, a priority and a domain.
Depending on the repackaging, this information can be \textbf{hidden} from or \textbf{exposed} to the network.

The hidden embedding tunnels a CF using an application layer protocol, which is the current state of the art.
Figure~\ref{fig:embeddingsomeip} shows an example that uses \ac{SOME/IP}. 
The CF ID is encoded in the message ID field and the data is embedded as payload. 
The example uses reserved UDP ports for \ac{SOME/IP} traffic. 
The Differentiated Services Code Point (DSCP) is populated with the CF priority.
The Ethernet header is generated by the network stack, which can be configured to map the DSCP to a \ac{PCP} for \ac{QoS} on Layer~2. 
All CFs in a domain are sent to the same IPv4 multicast address.

In contrast, embedding can happen in a completely exposed way. 
Here, the sender embeds the context of the CF only in packet header fields that are used for the forwarding decision in the network (Layer~2 to Layer~4). 
Figure~\ref{fig:embeddingl2} shows an example for exposed embedding in the Ethernet header.
For a discussion on the advantages of an embedding on the lowest possible layer see Section~\ref{subsec:network_design}.
The CF ID is encoded as a multicast destination MAC address.
Virtual LANs are created to separate bus domains.
The CF priority is mapped on the \ac{PCP} for Layer 2 \ac{QoS} differentiation.
We embed the message data and length into the Layer~2 payload.
The EtherType specifies the embedded data type, e.g., a custom type for embedded CAN data.

\vspace{-6pt}
\subsection{Separating In-Vehicle Control Flows in the Network}
\label{subsec:separation}
The embedded context information within the packet header fields as used for the forwarding decision is outlined in red in Figure~\ref{fig:embeddings}.  
A \textbf{Network Flow (NF)} is a sequence of contiguous packets of one or more CFs that have identical matchable header fields and are therefore treated equally by the network.
A NF is transported from a specific source to a destination in the network.
Using multicast, a NF can reach multiple sinks, which is a common use case in \acp{IVN}, since the same information is often needed at different \acp{ECU}.
The choice of the embedding approach is a key factor in distinguishing CFs.

With hidden embeddings, a \textbf{separation by domain} can be achieved if a domain identifier is encoded in the destination IP address. 
This creates a domain tunnel that forms a point-to-multipoint NF per sender and domain, which is the current state of the art.
The tunnel is identified by the source and destination IP addresses and UDP ports.
The NF destinations are the group of receivers of all CFs in the tunnel. 
The CFs of a domain cannot be differentiated by the network.

To improve the separation of CFs with hidden embeddings, a finer-grained \textbf{separation by topic} can be created.  
A topic can group a small number of related CFs, e.g., all communication for the engine or for lighting control.
In CAN bus architectures, each CF belongs to exactly one domain bus.
Introducing a new domain for CFs also means creating a new physical bus.
On an Ethernet backbone, a new topic requires only a new tunnel on an existing physical link.
As cross-domain communication increases, it seems advisable to form smaller groups of CFs than the original vehicle domains. 
The topic identifier can be encoded in the multicast IP address instead of the domain identifier so that each topic has a point-to-multipoint NF for each sender in the topic.

An exact \textbf{separation by message} type can only be achieved via exposed embedding. 
Each CF has its own point-to-multipoint NF, which can be identified and separated in the network by matching the multicast destination MAC address.

\subsection{Reliability and Security Considerations}
Reliable communication is essential for safety-critical traffic in the vehicle. 
Basic driving functionality must be guaranteed in order to achieve a safe state in an emergency, e.g., stopping at the edge of the road. 
A static configuration for safety-critical communication can be verified offline to ensure its correctness under all circumstances, and redundant paths can be confirmed to increase resilience.

The static flow and timing configuration is loaded in each forwarding device during boot, which reduces startup times because these flows are not set up via the controller. 
For changing this static configuration a firmware update is required as the \ac{SDN} controller can not alter it in any way. 
To achieve this, a protected separate flow table can be used that is always matched before dynamic rules. 
For the \ac{TSN} configuration, the priority queues can be partitioned in a set of static and dynamic queues~\cite{lbp-bptjr-21}. 
This allows the addition of new flows in dynamic priorities and protects the static priorities by design.
In the event of a controller failure or critical security incident, the static configuration serves as a fail-operational configuration.

Dynamic traffic that is not always needed or where the communication partners are not known before runtime can be controlled by the \ac{SDN} controller.
The controller can verify that the new communication is allowed and identify senders and receivers to create a precise flow that matches the header information from Layer~2 to Layer~4. 
In addition, bandwidth or time slots can be reserved for these flows.
An \ac{ACL} can define additional patterns that whitelist or blacklist dynamic flows. 
For example, complete protocols, such as ARP or ICMP, can be (dis-)allowed in the \ac{IVN} or dynamic communication can be blocked for some hosts. 

\vspace{-10pt}
\subsection{Impact on In-Vehicle Network Design}
\label{subsec:network_design}
When Control Flow (CF) information is embedded at the application layer, it cannot be used for forwarding decisions without violating the OSI layers. 
In \ac{SDN}, exposed embedding of the CF ID in any of the matched header fields from Layer 2 to 4 will enable a precise separation by message.
Still, there are several advantages of embeddings on the lowest possible layer.
Layer~2 information is only valid in the local \ac{IVN}, so Ethernet embeddings are not routable which can make attacks from outside the vehicle more difficult.
In addition, embeddings in the Ethernet header support the use of non-SDN switches with the same separation.
Embedding options are also affected by encryption, as the layers used for the forwarding decision are useless if encrypted.

When small control messages are embedded into Ethernet, their aggregation is an approach to reduce overhead. 
For example, a CAN message has a payload of 8 bytes, while an Ethernet frame has a minimum size of 64 bytes with a minimum payload of 42 bytes. 
Multiple CAN messages can be sent in the same frame to save bandwidth, but this also delays messages, increasing latency and jitter~\cite{krst-gsece-11}.
With different embedding approaches, aggregation can be hindered, e.g., exposed embeddings make aggregation impossible.

Control information is often transmitted in cycles with a similar data size.
\ac{TSN} schedules can be more efficient the better CFs can be distinguished in the network.
With exposed embeddings, timing can be determined more accurately, resulting in lower \ac{TDMA} reservation overhead.
On the other hand, schedule computation overhead increases as the number of small time slots increases.
Aggregation complicates timing computation due to varying sizes and intervals of packets.

\section{Validating the TSSDN Switching Architecture}
\label{sec:eval_validation}

\begin{figure}
  \centering
	\includegraphics[width=.85\linewidth, trim=0.62cm 0.62cm 0.62cm 0.62cm, clip=true]{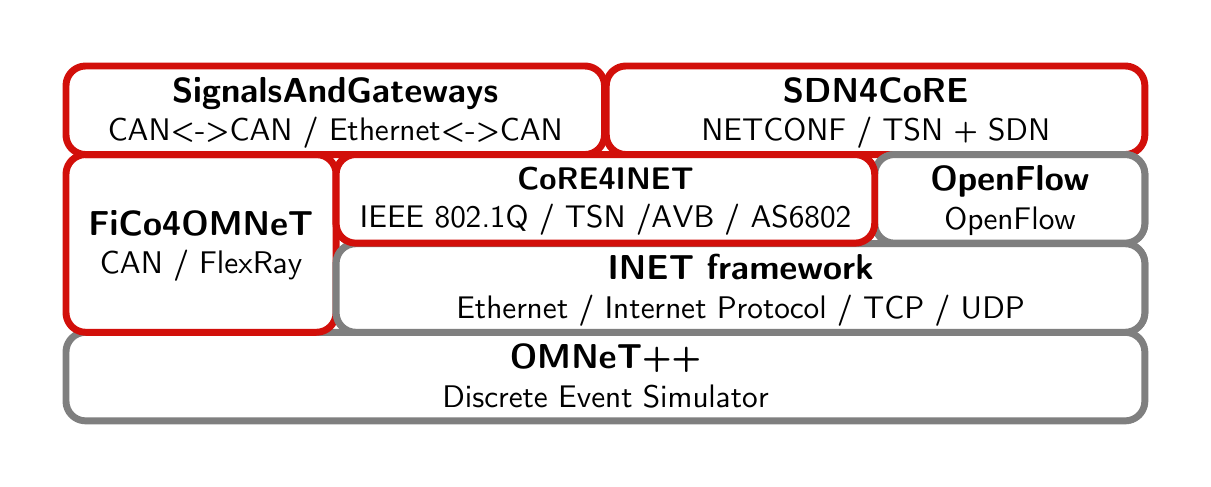}
	\caption{
    Simulation environment used for the evaluation available as open source, see \textit{github.com/CoRE-RG/SDN4CoRE}.
  }
	\label{fig:sim_env}
  \vspace{-12pt}
\end{figure}

The key performance characteristic of the proposed \ac{TSSDN} architecture is its seamless integration of SDN control with real-time communication. 
We evaluate this in a simulation environment (see Figure~\ref{fig:sim_env}) based on the OMNeT++ simulator~\cite{omnetpp}. 
Our model \ac{SDN4CoRE}~\cite{hmks-smsdn-19} implements the proposed architecture based on the INET framework~\cite{inet-framework}, the OpenFlowOMNeTSuite~\cite{kj-oeojr-13}, and our frameworks for \acp{IVN}~\cite{mkss-smcin-19}.
The models outlined in red are maintained by our research group and published as open source at \textit{\url{sim.core-rg.de}} and \textit{\url{github.com/CoRE-RG}}. 
We use a specific scenario with carefully designed TDMA traffic and varying amounts of cross traffic to analyze the impact of SDN control and schedule reconfiguration on real-time traffic.
This scenario is also available in the SDN4CoRE repository.

\vspace{-8pt}
\subsection{Network Topology and Traffic Configuration}
\begin{figure}
  \centering
	\includegraphics[width=0.95\linewidth, trim=0.62cm 0.62cm 0.62cm 0.62cm, clip=true]{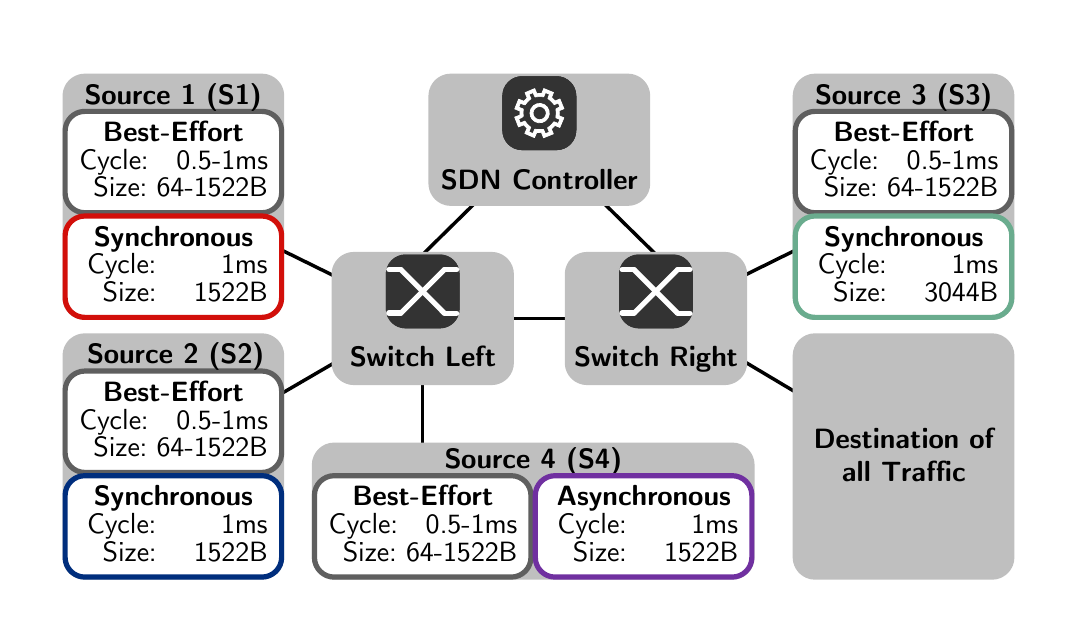}
	\caption{
    Evaluation network with two TSSDN switches, one SDN controller, four sources and one destination connected via \SI{100}{\mega\bit\per\second} links. 
    Sources 1 to 4 send best-effort, synchronous and asynchronous real-time traffic that differs in transmission cycle and Ethernet frame size.
  }
	\label{fig:case_study_topology}
  \vspace{-8pt}
\end{figure}

Figure~\ref{fig:case_study_topology} shows our network for the evaluation.
It consists of two \ac{TSSDN} switches, one \ac{SDN} controller, and four sources (S1-S4) sending to one destination.
All links have a bandwidth of \SI{100}{\mega\bit\per\second}.
With this setup we can analyze the timing of real-time communication under the load of cross traffic across multiple links and verify that timing requirements of \ac{TSN} are also met in \ac{TSSDN}.
For comparison, we consider the identical scenario with pure \ac{TSN} switches that perform stream reservation in a distributed manner and have a pre-configured \ac{TDMA} schedule, which is the current state of the art. 

All sources send BE traffic (\ac{PCP} 0) with varying cycle and frame sizes.
S1, S2, and S3 send synchronous traffic each at an individual priority (\ac{PCP} 5 to 7), which are scheduled in a \ac{TDMA} fashion with a period of \SI{1}{\milli\second}.
S1 and S2 send one maximum Ethernet frame per period and S3 sends two. 
S4 sends a maximum Ethernet frame every \SI{1}{\milli\second} as asynchronous traffic of medium priority (\ac{PCP} 4) directly after the \acf{SR} is completed.

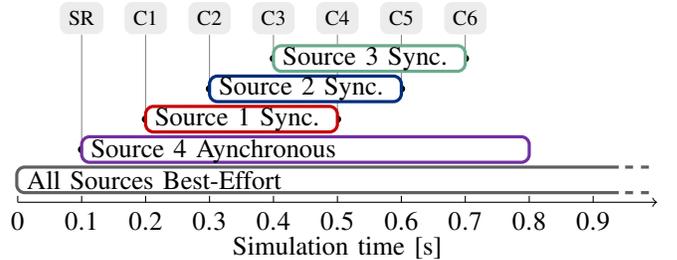
\begin{figure}
  \centering
	\begin{tikzpicture}[%
      scale=0.85,
    ]
    \draw[->] (0,0) -- (10,0);

    \node[anchor=north] at (0,0) {$0$};
    \draw (0,0.1) -- (0,0);
    \foreach \x in {1,...,9}{
        \draw (\x,0.1) -- (\x,0);
        \node[anchor=north] at (\x,0) {$0.\x$};
    }
    \node[anchor=north] at (5,-0.4) {$\text{Simulation time [s]}$};
    
    \node[small dot, pin={[pin distance=42pt]90:{SR}}] at (1,0.825) {};
    \node[small dot, pin={[pin distance=30.5pt]90:{C1}}] at (2,1.3) {};
    \node[small dot, pin={[pin distance=19pt]90:{C2}}] at (3,1.775) {};
    \node[small dot, pin={[pin distance=7.5pt]90:{C3}}] at (4,2.25) {};
    \node[small dot, pin={[pin distance=30.5pt]90:{C4}}] at (5,1.3) {};
    \node[small dot, pin={[pin distance=19pt]90:{C5}}] at (6,1.775) {};
    \node[small dot, pin={[pin distance=7.5pt]90:{C6}}] at (7,2.25) {};

    \draw[CoreDarkGray, very thick,rounded corners=3pt]
                (9.4,0.15) -- (-0.015,0.15) -- (-0.015,0.55) -- (9.4,0.55);
    \draw[CoreDarkGray, very thick, dashed] (9.5,0.15) -- (9.9,0.15);
    \draw[CoreDarkGray, very thick, dashed] (9.5,0.55) -- (9.9,0.55);
    \node[anchor=west] at (0,0.35) {$\text{All Sources Best-Effort}$};
    \draw[CoreMagenta, very thick,rounded corners=3pt,fill=white] (1,0.625) rectangle (8,1.025);    
    \node[anchor=west] at (1,0.8) {$\text{Source 4 Aynchronous}$};
    \draw[CoreRed, very thick,rounded corners=3pt,fill=white] (2,1.1) rectangle (5,1.5);
    \node[anchor=west] at (2,1.3) {$\text{Source 1 Sync.}$};
    \draw[CoreBlue, very thick,rounded corners=3pt,fill=white] (3,1.575) rectangle (6,1.975);
    \node[anchor=west] at (3,1.775) {$\text{Source 2 Sync.}$};
    \draw[CoreGreen, very thick,rounded corners=3pt,fill=white] (4,2.05) rectangle (7,2.45);
    \node[anchor=west] at (4,2.25) {$\text{Source 3 Sync.}$};
  \end{tikzpicture}
	\caption{
    Timeline indicating the start and stop times of traffic with their stream reservation (SR) or configurations (C1--C6).
  }
	\label{fig:case_study_timeline}
\end{figure}

The timeline in Figure~\ref{fig:case_study_timeline} visualizes the start and stop times of real-time communication.
For asynchronous traffic, the controller performs the \ac{SR} using OpenFlow.
Prior to changes in synchronous traffic, we assume that the \ac{SDN} controller has been informed  and the \acf{GCL} schedule is reconfigured (C1 -- C6) via NETCONF in the period before the first or after the last frame of the altering flow.
In the pure TSN version, the \ac{GCL} schedule is statically configured and the switches perform the \ac{SRP} of \ac{TSN} independently.

Our scenario contains all basic operations for reconfiguring the network, which follows from the changes in synchronous real-time communication.
Synchronous traffic is added to an empty schedule (C1) and to an existing schedule (C2).
The time slot for S3 is inserted between the time slots for S1 and S2 (C3), so that the time slot for S2 needs shifting to an earlier time and for S1 to a later time in the period.
This allows us to analyze the impact of combining multiple basic operations with different mandatory commit sequences in one transaction.
Each synchronous transmitter stops after sending for \SI{300}{\milli\second}.
The network removes time slots from existing schedules (C4, C5) and returns to an empty schedule at the end (C6).

\subsection{TDMA Schedule and Worst-Case Analysis}
\label{subsec:schedule}

\begin{figure}
  \begin{minipage}{\linewidth}
  
    \begin{tikzpicture}[scale=0.65]
      \node[anchor=north west] at (-3.2,0.6) {Source 1:};
      \fill[LightCoreGray, very thick, rounded corners=3pt] (0,0) rectangle (10,0.5);
      \fill[CoreBlack, very thick, rounded corners=3pt] (3.25,0) rectangle (4.5,0.5);
      \node[anchor=north west,text=white] at (3.25,0.6) {GB};
      \fill[CoreRed, very thick, rounded corners=3pt] (4.50,0) rectangle (5.75,0.5);
      \node[anchor=north west,text=white] at (4.50,0.6) {S1};
    \end{tikzpicture}  
  
    \vspace{-2pt}
  
    \begin{tikzpicture}[scale=0.65]
      \node[anchor=north west] at (-3.2,0.6) {Source 2:};
      \fill[LightCoreGray, very thick, rounded corners=3pt] (0,0) rectangle (10,0.5);
      \fill[CoreBlack, very thick, rounded corners=3pt] (1.25,0) rectangle (2.5,0.5);
      \node[anchor=north west,text=white] at (1.25,0.6) {GB};
      \fill[CoreBlue, very thick, rounded corners=3pt] (2.50,0) rectangle (3.75,0.5);
      \node[anchor=north west,text=white] at (2.50,0.6) {S2};
    \end{tikzpicture}  
  
    \vspace{-2pt}
  
    \begin{tikzpicture}[scale=0.65]
      \node[anchor=north west] at (-3.2,0.6) {Source 3:};
      \fill[LightCoreGray, very thick, rounded corners=3pt] (0,0) rectangle (10,0.5);
      \fill[CoreBlack, very thick, rounded corners=3pt] (3.75,0) rectangle (5,0.5);
      \node[anchor=north west,text=white] at (3.75,0.6) {GB};
      \fill[CoreGreen, very thick, rounded corners=3pt] (5,0) rectangle (7.5,0.5);
      \node[anchor=north west,text=white] at (5,0.6) {S3};
    \end{tikzpicture}  
  
    \vspace{-1pt}
  
    \begin{tikzpicture}[scale=0.65]    
      \node[anchor=north west] at (-3.2,0) { };
      \draw[very thick] (0,-0.3) -- (10,-0.3);
      \draw[very thick] (0,-0.2) -- (0,-0.4);
      \node[anchor=north] at (0,-0.3) {$0$};
      \draw[very thick] (10,-0.2) -- (10,-0.4);
      \node[anchor=north] at (10,-0.3) {$1$};
      \foreach \x in {1,...,9}{
          \draw (\x,-0.2) -- (\x,-0.4);
          \node[anchor=north] at (\x,-0.3) {$0.\x$};
      }
      \node[anchor=north] at (5,-0.9) {Time slots in the GCL period [\si[]{\milli\second}]};
  
    \end{tikzpicture}
  
    \vspace{-5pt}
  
    \caption{
      Transmission windows of the synchronous sources in the period of the \acf{GCL} schedule.
    }
    \label{fig:sendwindows}
    
  \end{minipage}
  
  \vspace{7pt}
  
  \begin{minipage}{\linewidth}
    \begin{tikzpicture}[scale=0.65]
      \node[anchor=north west] at (-3.2,0.6) {\textbf{\SIrange{0}{0.2}{\second}: Initial State}};
    \end{tikzpicture}
    
    \vspace{-3pt}
    
    \begin{tikzpicture}[scale=0.65]
      \node[anchor=north west] at (-3.2,0.6) {Switch Left};
      \fill[LightCoreGray, very thick, rounded corners=3pt] (0,0) rectangle (10,0.5);
      \node[anchor=north west] at (0,0.65) {$\text{All Open}$};
    \end{tikzpicture}
    
    \vspace{-4pt}
    
    \begin{tikzpicture}[scale=0.65]
      \node[anchor=north west] at (-3.2,0.6) {Switch Right};
      \fill[LightCoreGray, very thick, rounded corners=3pt] (0,0) rectangle (10,0.5);
      \node[anchor=north west] at (0,0.65) {$\text{All Open}$};
    \end{tikzpicture}
    
    \vspace{-1pt}
  
    \begin{tikzpicture}[scale=0.65]
      \node[anchor=north west] at (-3.2,0.6) {\textbf{\SIrange{0.2}{0.3}{\second}: C1 add Source 1 Synchronous}};
    \end{tikzpicture}  
  
    \vspace{-3pt}  
  
    \begin{tikzpicture}[scale=0.65]
      \node[anchor=north west] at (-3.2,0.6) {Switch Left};
      \fill[LightCoreGray, very thick, rounded corners=3pt] (0,0) rectangle (10,0.5);
      \fill[CoreBlack, very thick, rounded corners=3pt] (4.5,0) rectangle (5.74,0.5);
      \node[anchor=north west,text=white] at (4.5,0.6) {GB};
      \fill[CoreRed, very thick, rounded corners=3pt] (5.74,0) rectangle (6.97,0.5);
      \node[anchor=north west,text=white] at (5.74,0.6) {S1};
    \end{tikzpicture}  
  
    \vspace{-2pt}
  
    \begin{tikzpicture}[scale=0.65]
      \node[anchor=north west] at (-3.2,0.6) {Switch Right};
      \fill[LightCoreGray, very thick, rounded corners=3pt] (0,0) rectangle (10,0.5);
      \fill[CoreBlack, very thick, rounded corners=3pt] (5.75,0) rectangle (6.99,0.5);
      \node[anchor=north west,text=white] at (5.75,0.6) {GB};
      \fill[CoreRed, very thick, rounded corners=3pt] (6.99,0) rectangle (8.23,0.5);
      \node[anchor=north west,text=white] at (6.99,0.6) {S1};
    \end{tikzpicture}
    
    \vspace{-1pt}
  
    \begin{tikzpicture}[scale=0.65]
      \node[anchor=north west] at (-3.2,0.6) {\textbf{\SIrange{0.3}{0.4}{\second}: C2 add Source 2 Synchronous}};
    \end{tikzpicture}  
  
    \vspace{-3pt}  
  
    \begin{tikzpicture}[scale=0.65]
      \node[anchor=north west] at (-3.2,0.6) {Switch Left};
      \fill[LightCoreGray, very thick, rounded corners=3pt] (0,0) rectangle (10,0.5);
      \fill[CoreBlack, very thick, rounded corners=3pt] (3.25,0) rectangle (4.49,0.5);
      \node[anchor=north west,text=white] at (3.25,0.6) {GB};
      \fill[CoreBlue, very thick, rounded corners=3pt] (4.49,0) rectangle (5.74,0.5);
      \node[anchor=north west,text=white] at (4.49,0.6) {S2};
      \fill[CoreRed, very thick, rounded corners=3pt] (5.74,0) rectangle (6.97,0.5);
      \node[anchor=north west,text=white] at (5.74,0.6) {S1};
    \end{tikzpicture}  
  
    \vspace{-2pt}
    
    \begin{tikzpicture}[scale=0.65]
      \node[anchor=north west] at (-3.2,0.6) {Switch Right};
      \fill[LightCoreGray, very thick, rounded corners=3pt] (0,0) rectangle (10,0.5);
      \fill[CoreBlack, very thick, rounded corners=3pt] (4.50,0) rectangle (5.74,0.5);
      \node[anchor=north west,text=white] at (4.50,0.6) {GB};
      \fill[CoreBlue, very thick, rounded corners=3pt] (5.74,0) rectangle (6.99,0.5);
      \node[anchor=north west,text=white] at (5.74,0.6) {S2};
      \fill[CoreRed, very thick, rounded corners=3pt] (6.99,0) rectangle (8.23,0.5);
      \node[anchor=north west,text=white] at (6.99,0.6) {S1};
    \end{tikzpicture}
    
    \vspace{-1pt}
  
    \begin{tikzpicture}[scale=0.65]
      \node[anchor=north west] at (-3.2,0.6) {\textbf{\SIrange{0.4}{0.5}{\second}: C3 add Source 3 Synchronous}};
      \node[anchor=north west] at (-3.2,0) {(TSN static configuration)};
    \end{tikzpicture}  
  
    \vspace{-3pt}  
  
    \begin{tikzpicture}[scale=0.65]
      \node[anchor=north west] at (-3.2,0.6) {Switch Left};
      \fill[LightCoreGray, very thick, rounded corners=3pt] (0,0) rectangle (10,0.5);
      \fill[CoreBlack, very thick, rounded corners=3pt] (2.5,0) rectangle (3.75,0.5);
      \node[anchor=north west,text=white] at (2.5,0.6) {GB};
      \fill[CoreBlue, very thick, rounded corners=3pt] (3.75,0) rectangle (4.97,0.5);
      \node[anchor=north west,text=white] at (3.75,0.6) {S2};
      \fill[CoreBlack, very thick, rounded corners=3pt] (6.25,0) rectangle (7.5,0.5);
      \node[anchor=north west,text=white] at (6.25,0.6) {GB};
      \fill[CoreRed, very thick, rounded corners=3pt] (7.5,0) rectangle (8.72,0.5);
      \node[anchor=north west,text=white] at (7.5,0.6) {S1};
    \end{tikzpicture}  
  
    \vspace{-2pt}
    
    \begin{tikzpicture}[scale=0.65]
      \node[anchor=north west] at (-3.2,0.6) {Switch Right};
      \fill[LightCoreGray, very thick, rounded corners=3pt] (0,0) rectangle (10,0.5);
      \fill[CoreBlack, very thick, rounded corners=3pt] (3.75,0) rectangle (4.99,0.5);
      \node[anchor=north west,text=white] at (3.75,0.6) {GB};
      \fill[CoreBlue, very thick, rounded corners=3pt] (4.99,0) rectangle (6.24,0.5);
      \node[anchor=north west,text=white] at (4.99,0.6) {S2};
      \fill[CoreGreen, very thick, rounded corners=3pt] (6.24,0) rectangle (8.74,0.5);
      \node[anchor=north west,text=white] at (6.24,0.6) {S3};
      \fill[CoreRed, very thick, rounded corners=3pt] (8.74,0) rectangle (9.90,0.5);
      \node[anchor=north west,text=white] at (8.74,0.6) {S1};
    \end{tikzpicture}
    
    \vspace{-1pt}
  
    \begin{tikzpicture}[scale=0.65]
      \node[anchor=north west] at (-3.2,0.6) {\textbf{\SIrange{0.5}{0.6}{\second}: C4 remove Source 1 Synchronous}};
    \end{tikzpicture}  
  
    \vspace{-3pt}  
  
    \begin{tikzpicture}[scale=0.65]
      \node[anchor=north west] at (-3.2,0.6) {Switch Left};
      \fill[LightCoreGray, very thick, rounded corners=3pt] (0,0) rectangle (10,0.5);
      \fill[CoreBlack, very thick, rounded corners=3pt] (2.5,0) rectangle (3.75,0.5);
      \node[anchor=north west,text=white] at (2.5,0.6) {GB};
      \fill[CoreBlue, very thick, rounded corners=3pt] (3.75,0) rectangle (4.97,0.5);
      \node[anchor=north west,text=white] at (3.75,0.6) {S2};
    \end{tikzpicture}  
  
    \vspace{-2pt}
    
    \begin{tikzpicture}[scale=0.65]
      \node[anchor=north west] at (-3.2,0.6) {Switch Right};
      \fill[LightCoreGray, very thick, rounded corners=3pt] (0,0) rectangle (10,0.5);
      \fill[CoreBlack, very thick, rounded corners=3pt] (3.75,0) rectangle (4.99,0.5);
      \node[anchor=north west,text=white] at (3.75,0.6) {GB};
      \fill[CoreBlue, very thick, rounded corners=3pt] (4.99,0) rectangle (6.24,0.5);
      \node[anchor=north west,text=white] at (4.99,0.6) {S2};
      \fill[CoreGreen, very thick, rounded corners=3pt] (6.24,0) rectangle (8.74,0.5);
      \node[anchor=north west,text=white] at (6.24,0.6) {S3};
    \end{tikzpicture}
    
    \vspace{-1pt}
  
    \begin{tikzpicture}[scale=0.65]
      \node[anchor=north west] at (-3.2,0.6) {\textbf{\SIrange{0.6}{0.7}{\second}: C5 remove Source 2 Synchronous}};
    \end{tikzpicture}  
  
    \vspace{-3pt}  
  
    \begin{tikzpicture}[scale=0.65]
      \node[anchor=north west] at (-3.2,0.6) {Switch Left};
      \fill[LightCoreGray, very thick, rounded corners=3pt] (0,0) rectangle (10,0.5);
    \end{tikzpicture}  
  
    \vspace{-2pt}
    
    \begin{tikzpicture}[scale=0.65]
      \node[anchor=north west] at (-3.2,0.6) {Switch Right};
      \fill[LightCoreGray, very thick, rounded corners=3pt] (0,0) rectangle (10,0.5);
      \fill[CoreBlack, very thick, rounded corners=3pt] (4.99,0) rectangle (6.24,0.5);
      \node[anchor=north west,text=white] at (4.99,0.6) {GB};
      \fill[CoreGreen, very thick, rounded corners=3pt] (6.24,0) rectangle (8.74,0.5);
      \node[anchor=north west,text=white] at (6.24,0.6) {S3};
    \end{tikzpicture}
  
    \vspace{-1pt}
  
    \begin{tikzpicture}[scale=0.65]
      \node[anchor=north west] at (-3.2,0.6) {\textbf{\SI{0.7}{\second} to $\infty$: C6 remove Source 3 Synchronous}};
    \end{tikzpicture}
    
    \vspace{-3pt}
    
    \begin{tikzpicture}[scale=0.65]
      \node[anchor=north west] at (-3.2,0.6) {Switch Left};
      \fill[LightCoreGray, very thick, rounded corners=3pt] (0,0) rectangle (10,0.5);
    \end{tikzpicture}
    
    \vspace{-2pt}
    
    \begin{tikzpicture}[scale=0.65]
      \node[anchor=north west] at (-3.2,0.6) {Switch Right};
      \fill[LightCoreGray, very thick, rounded corners=3pt] (0,0) rectangle (10,0.5);
    \end{tikzpicture}
    
    \vspace{-1pt}
  
    \begin{tikzpicture}[scale=0.65]    
      \node[anchor=north west] at (-3.2,0) { };
      \draw[very thick] (0,-0.3) -- (10,-0.3);
      \draw[very thick] (0,-0.2) -- (0,-0.4);
      \node[anchor=north] at (0,-0.3) {$0$};
      \draw[very thick] (10,-0.2) -- (10,-0.4);
      \node[anchor=north] at (10,-0.3) {$1$};
      \foreach \x in {1,...,9}{
          \draw (\x,-0.2) -- (\x,-0.4);
          \node[anchor=north] at (\x,-0.3) {$0.\x$};
      }
      \node[anchor=north] at (5,-0.9) {Time slots in the GCL period [\si[]{\milli\second}]};
    \end{tikzpicture}
  
    \vspace{-1pt}
  
  \caption{
    \acf{GCL} schedule of the two switches, which is reconfigured according to the changes in synchronous traffic (C1 to C6).
    The time slots for S1, S2 and S3 are scheduled along their path.
    Guard Bands (GB) with all gates closed are added to prevent delays from competing traffic.
    The TSN-only version is statically configured as depicted in C3. 
  }
  \label{fig:controllist}
  \end{minipage}
  \vspace{-10pt}
\end{figure}

We determine periodic \ac{TDMA} schedules that coordinate synchronous flows in the \SI{1}{\milli\second} period. 
The transmission windows of synchronous traffic at each source are shown  in Figure~\ref{fig:sendwindows} for the entire period.
Figure~\ref{fig:controllist} displays the \ac{GCL} schedule of the two switches.
The pure TSN network is pre-configured on startup with configuration C3, which includes all synchronous flows.
In the TSSDN variant of the network, the \ac{GCL} configuration is updated according to the active synchronous traffic (C1 to C6).
At the beginning, all gates are open because there is no synchronous traffic yet. 

For each sender, our \ac{GCL} configuration exclusively assigns one time slot to its 802.1Q priority. 
A Guard Band (GB), in which all gates are closed, is added to ensure that synchronous traffic cannot be delayed by other traffic.
All gates are closed so that the transmission time of a maximum Ethernet frame ($\SI{1522}{\byte} + \SI{8}{\byte} \text{ preamble}$) can be completed.
This takes \SI{123.36}{\micro\second} on a \SI{100}{\mega\bit\per\second} link, including the inter frame gap ($T_{\text{ifg}}=\SI{0.96}{\micro\second}$). 
To ensure timely transmission, the \acp{GCL} at the senders are also scheduled with a GB before their transmission windows (see Figure~\ref{fig:sendwindows}).

Synchronous traffic, which is precisely scheduled, cannot be delayed as it is protected by a GB. 
Its maximal end-to-end latency $T_{\text{L}}$ can be analytically calculated  equivalent to the best-case latency when all clocks are perfectly synchronized.
In our evaluation, we take a closer look at the latency of S1. 
The calculations for S2 and S3 are not shown in detail.

Reconfiguration C1 adds the synchronous flow from S1 at \SI{0.2}{\second} simulation time.
S1 transmits one full size frame \SI{450}{\micro\second} after the start of each period. 
A time slot is reserved along the path.
After the delay for transmission ($T_{\text{trans}}=\SI{122.4}{\micro\second}$) and forwarding ($T_{\text{fwd}}=\SI{3}{\micro\second}$) the frame arrives at the output of Switch Left at \SI{575}{\micro\second} and its time slot begins.
A GB is inserted before it.
All gates are opened again when the frame completes transmission at \SI{700}{\micro\second} and its time slot at Switch Right begins. 
It ends at \SI{825}{\micro\second} and the \ac{GCL} is configured accordingly.
The analytic end-to-end latency bound for synchronous traffic from S1 with configuration C1 ($T^{\text{S1}}_{\text{L-C1}}$) reads:
\begin{equation}
  \begin{aligned}
    T^{\text{S1}}_{\text{L-C1}} &= 3 * T_{\text{trans}} + 2 * T_{\text{fwd}}
    \\
    & = 3 * \SI{122.4}{\micro\second} + 2 * \SI{3}{\micro\second} = \SI{373.2}{\micro\second}
  \end{aligned}
  \label{eq:wc_s1_sync}
\end{equation}

The controller activates C2 when the synchronous traffic from S2 starts at \SI{0.3}{\second} simulation time.
S2 sends a full size frame \SI{250}{\micro\second} after the start of each period.
Thus, traffic from S2 arrives at \SI{375}{\micro\second} at the output of Switch Left.
To save bandwidth the time slot of S2 is scheduled 
to finish exactly before the time slot of S1. 
The GB is shifted accordingly.
This introduces a queueing delay of \SI{75}{\micro\second}.
The frame arrives at Switch Right at the beginning of its time slot to be transmitted. 
This results in a latency bound of $T^{\text{S2}}_{\text{L-C2}}=\SI{448.2}{\micro\second}$.

The schedule is reconfigured for S3 at \SI{0.4}{\second} simulation time (C3). 
S3 sends two full size frames \SI{500}{\micro\second} after the start of each period.
This new schedule minimizes the latency of S3 ($T^{\text{S3}}_{\text{L-C3}}=\SI{247.8}{\micro\second}$) by adding a time slot on Switch Right from \SI{625}{\micro\second} until \SI{875}{\micro\second}.
This requires to shift the time slots of S1 to a later time and S2 to an earlier time, which also changes their latency bound.
The time slot for S2 is now \SI{75}{\micro\second} earlier on both switches which also makes up for the queueing delay and reduces the latency bound to $T^{\text{S2}}_{\text{L-C3}}=\SI{373.2}{\micro\second}$.
The time slot for S1 is delayed by \SI{175}{\micro\second} which introduces a queuing delay ($T_{\text{queue}}$) and changes the latency bound to
\begin{equation}
  \begin{aligned}
    T^{\text{S1}}_{\text{L-C3}} &= 3 * T_{\text{trans}} + 2 * T_{\text{fwd}} + T_{\text{queue}}
    \\
    & = 3 * \SI{122.4}{\micro\second} + 2 * \SI{3}{\micro\second} + \SI{175}{\micro\second}= \SI{548.2}{\micro\second}
  \end{aligned}
  \label{eq:wc_s1_c3_sync}
\end{equation}

Asynchronous traffic can be delayed by cross traffic and the \ac{TSN} gates.
We examined the coexistence of scheduled and bandwidth reserved traffic and the impact of cross traffic in previous work~\cite{slksh-bhcan-15, msks-eatts-13}.
In the worst-case, the asynchronous traffic from S4 is delayed by one full size \ac{BE} frame ($T_{\text{BE max}}$) and the inter frame gap ($T_{\text{ifg}}$) on every hop.
In our scenario, it can also be delayed through the schedule at each port by the duration of the GB and the time slots of synchronous traffic. 
The schedules for Switch Left and Right are aligned so that a frame cannot be delayed on both hops.
For the configuration C3 this results in a maximum interference through the schedule of $T_{\text{mi}} = \SI{616.8}{\micro\second}$.
The analytical end-to-end latency bound for the asynchronous traffic from S4 ($T^{\text{S4}}_{\text{wc}}$) reads: 
\setlength\arraycolsep{1.4pt}
\begin{eqnarray}
  T^{\text{S4}}_{\text{L}_{\text{wc}}} &=& T_{\text{mi}} + 3 * (T_{\text{BE max}} + T_{\text{ifg}} + T_{\text{trans}}) + 2 * T_{\text{fwd}}
    \\
    &=& \SI{616.8}{\micro\second} + 3 * (\SI{122.4}{\micro\second} + \SI{0.96}{\micro\second} + \SI{122.4}{\micro\second}) + 2 * \SI{3}{\micro\second}  \nonumber
    \\
    &=& \SI{1.36}{\milli\second} \nonumber
    \label{eq:wc_s2_async}
\end{eqnarray}

\vspace{-10pt}
\subsection{Impact of SDN on Real-Time Flows}
Figure~\ref{fig:latencyallminmax} shows the minimum, maximum and average end-to-end latency for one exemplary flow of each traffic class from \SI{1}{\second} simulation runs. 
To ensure that the data is not affected by the timing of traffic in the period, the minimum, maximum, and average from simulations with 20 different seeds for the start time of the asynchronous traffic are shown.
The queues in our network are infinite so all packets will arrive at there destination eventually and no packets are lost.

All traffic shows identical or lower end-to-end latency in the \ac{TSSDN} variant than in pure TSN. 
This is due to the re-scheduled time slots (C1 to C6) that match the synchronous traffic exactly without surplus bandwidth.
Given the same configuration for both variants (after C3), the synchronous traffic of S1 has the same constant latency corresponding to the analytical bound (see Equation~\ref{eq:wc_s1_c3_sync}).
Before C3 is applied, S1 has a smaller latency in \ac{TSSDN} as calculated in Equation~\ref{eq:wc_s1_sync}.

S4 starts sending asynchronous traffic after the \acf{SR} at \SI{0.1}{\second}.
Again, latency is lower with the TSSDN variant because less bandwidth is reserved for the scheduled traffic.
The latency varies largely for the 20 different seeds for the start time in the period, which is to be expected since the frame delays fluctuate by the schedule.  
Still, the maximum latency never exceeds the analytic bound of Equation~\ref{eq:wc_s2_async}. 

The BE traffic flows without  worst-case guaranties.
Frame sizes vary between minimal and maximal Ethernet frames, which explains the small minimum latency.
In TSSDN, \ac{BE} flows experiences a larger delay if set up by the controller.

The latency for flow installation is avoided for asynchronous traffic with the \acf{SRP}.
As such flows are already installed in the \ac{TSSDN} switches during the \ac{SR}, no further inspection by the \ac{SDN} controller is needed.
Thus, additional latency ceases for \ac{TSSDN} after the \ac{SR}.

Table~\ref{tab:sr_duration} compares the SR duration of TSN and TSSDN, which vary largely with the 20 seeds (uniformly distributed over the \SI{1}{\milli\second} period of the schedule).
The TSSDN variant has a lower maximum delay when set up at simulation time \SI{0.1}{\second}, since no gate control schedule is active at this time.
We also ran the simulation in TSSDN starting the SR at \SI{0.45}{\second} with configuration C3 active and thus the same schedule as in the TSN-only variant.
The results show an additional delay in TSSDN caused by the communication between the switches and the controller.
This is in agreement with our previous findings for SR in TSSDN~\cite{hmks-snsti-19} but extends insights to the impact of scheduled traffic on SR sent as BE traffic in both cases, TSSDN and plain TSN.

\vspace{-10pt}
\subsection{Impact of Transactional Updates}
The impact of the synchronous and ordered update methods on the latency of the scheduled traffic is shown in Figure~\ref{fig:tssdnlatency}.
We vary the commit execution time to find the worst case for each configuration method.
For simple unordered reconfiguration, the result is completely random and not shown.
For synchronous updates in both switches, the latency between two configurations is constant for each interval and matches the calculated bounds obtained in Section~\ref{subsec:schedule}.
After C3, the latency of S1 and S2 changes as time slots are shifted for the additional flow from S3.

\begin{figure}
    \centering
    \captionof{table}{\acf{SR} delay of TSN and TSSDN w/ and w/o Gate Control (GC) for 20 different SR start times. 
    }
    \label{tab:sr_duration}
    \setlength{\tabcolsep}{2pt}
    \footnotesize
    \begin{tabularx}{\linewidth}{l l X c c c}
        \toprule
        \textbf{Variant} && \textbf{SR start} & \multicolumn{3}{c}{\textbf{SR duration}}       
        \\ 
        && (20 seeds; \SI{0.05}{\milli\second} steps) & Minimum & Average & Maximum 
        \\ \midrule
        \textbf{TSN} & w/ GC & \SIrange{0.1}{0.10095}{\second} & \SI{0.115}{\milli\second} & \SI{0.775}{\milli\second} & \SI{1.404}{\milli\second}
        \\
        \textbf{TSSDN} & w/o GC & \SIrange{0.1}{0.10095}{\second} & \SI{0.828}{\milli\second} & \SI{0.862}{\milli\second} & \SI{1.001}{\milli\second}
        \\
        \textbf{TSSDN} & w/ GC & \SIrange{0.45}{0.45095}{\second} & \SI{0.828}{\milli\second} & \SI{1.495}{\milli\second} & \SI{2.370}{\milli\second}
        \\
        \bottomrule
    \end{tabularx}
    \medskip
    \vspace{2pt}
    
    \normalsize
    \begin{tikzpicture}
      \begin{groupplot}[
          width = .99\linewidth, height = 0.38\linewidth,
          group style = {
            group size = 1 by 3,
            horizontal sep = 0pt, vertical sep = 0pt,
            xlabels at = edge bottom,
            xticklabels at = edge bottom, 
            ylabels at = edge left,
            yticklabels at = edge left, 
            },
          xmajorgrids=true,
          change x base,
          x SI prefix=milli, x unit=\milli\second,
          xlabel = {End-to-end latency},
          xmin = 0, xmax = 0.003,
          clip=true,   
          symbolic y coords={
            Min0,Min,Min1,
            Sep11,Sep12,Sep13,
            Avg0,Avg,Avg1,
            Sep21,Sep22,Sep23,
            Max0,Max,Max1
            },
          ytick={Min, Avg, Max},
          ytick style = {draw=none},
          legend style={
            font=\small,
            at={(0.98,0.65)}, 
            legend cell align={left},
            anchor=east, 
            legend columns=1, 
            },
          ]
        \nextgroupplot[ylabel = {S1 Sync},
          extra x ticks={0.0003732,0.0005485},  
          extra x tick labels={}, 
          extra x tick style={grid=major,major grid style={thick,draw=black}}
          ] 
          \addplot [minmax][CoreRed,forget plot,only marks]  table [col sep=comma]{./results/max_latency/exp_tsn_s1_sync.csv};
          \addplot [avg][CoreRed,only marks, mark=*] table [col sep=comma]{./results/max_latency/exp_tsn_s1_sync.csv};
          \addplot [minmax][CoreBlue,forget plot,only marks]  table [col sep=comma]{./results/max_latency/exp_tssdn_s1_sync_from400.csv};
          \addplot [avg][CoreBlue,only marks, mark=square*] table [col sep=comma]{./results/max_latency/exp_tssdn_s1_sync_from400.csv};
          \addplot [minmax][CoreBlue,forget plot,only marks]  table [col sep=comma]{./results/max_latency/exp_tssdn_s1_sync_until400.csv};
          \addplot [avg][CoreBlue,only marks, mark=square*, fill=white] table [col sep=comma]{./results/max_latency/exp_tssdn_s1_sync_until400.csv};
  
          \addlegendimage{legend image with text = \textbf{|}}
          \legend{TSN, TSSDN, TSSDN before C3 , Analytical bound }; 
  
        \nextgroupplot[ylabel = {S4 Async},
          extra x ticks={0.0013601}, 
          extra x tick labels={}, 
          extra x tick style={grid=major,major grid style={thick,draw=black}}] 
          \addplot [minmax][CoreRed,forget plot,only marks]  table [col sep=comma]{./results/max_latency/exp_tsn_s4_async.csv};
          \addplot [avg][CoreRed,only marks, mark=*] table [col sep=comma]{./results/max_latency/exp_tsn_s4_async.csv};
          \addplot [minmax][CoreBlue,forget plot,only marks]  table [col sep=comma]{./results/max_latency/exp_tssdn_s4_async.csv};
          \addplot [avg][CoreBlue,only marks, mark=square*] table [col sep=comma]{./results/max_latency/exp_tssdn_s4_async.csv};
  
        \nextgroupplot[ylabel = {S4 BE},]
          \addplot [minmax][CoreRed,forget plot,only marks]  table [col sep=comma]{./results/max_latency/exp_tsn_s4_ct.csv};
          \addplot [avg][CoreRed,only marks, mark=*] table [col sep=comma]{./results/max_latency/exp_tsn_s4_ct.csv};
          \addplot [minmax][CoreBlue,forget plot,only marks]  table [col sep=comma]{./results/max_latency/exp_tssdn_s4_ct.csv};
          \addplot [avg][CoreBlue,only marks, mark=square*] table [col sep=comma]{./results/max_latency/exp_tssdn_s4_ct.csv};
         
      \end{groupplot}
      
      \node[small dot,pin={[pin distance=14pt]105:{$T^{\text{S1}}_{\text{L-C1}}=$ \SI{0.37}{\milli\second}}}] at (0.892,1.2) {};
      \node[small dot,pin={[pin distance=8pt]-30:{$T^{\text{S1}}_{\text{L-C3}}=$ \SI{0.55}{\milli\second}}}] at (1.315,1.2) {};
      \node[small dot,pin={[pin distance=8pt]-30:{$T^{\text{S2}}_{\text{L}_{\text{wc}}}=$ \SI{1.36}{\milli\second}}}] at (3.257,-0.93) {};
    \end{tikzpicture}
    \captionof{figure}{
      End-to-end latency comparison for the three traffic classes in TSN and TSSDN.
      The minimum, average, maximum, and the analytical bound are depicted.
      For each min/avg/max, again the average and deviation for minimum and maximum is indicated from simulations with 20 seeds.
    }
    \label{fig:latencyallminmax}
  \medskip
  \vspace{-2pt}
  
    \begin{tikzpicture}[]
      \tikzset{
        pin anchor/.style={tikz@pin@post/.append style={anchor=#1}},
        label anchor/.style={tikz@label@post/.append style={anchor=#1}},
      }
      \begin{groupplot}[
        width=\linewidth, height=.6\linewidth,
        group style = {
          group size = 1 by 3,
          horizontal sep = 0pt, vertical sep = 0pt,
          xlabels at = edge bottom,
          xticklabels at = edge bottom, 
          ylabels at = edge left,
          yticklabels at = edge left, 
          },
        legend pos=north east,
        legend style={
            at={(0.97,0.75)}, 
            legend cell align={left},
            anchor=east, 
            legend columns=1, 
            font=\small
          },
        clip=true,   
        x unit=\second,
        change y base,
        y SI prefix=milli,
        xlabel=Simulation time,
        yticklabel pos=left,
        xmajorgrids=true,
        xmin=0.15,
        xmax=0.75,
        ymin=0,
        ymax=0.0018,
        ytick={0.0002,0.0004,0.0006,0.0008,0.0010,0.0012,0.0014,0.0016},
        every axis title/.style={at={(0.01,1)}, right, yshift=-10pt},
        ]
  
      \nextgroupplot[title={Synchronous update}, height=.39\linewidth, ymax=0.00099]
        \addplot[ CoreRed,
          very thick,
          ] table [x, y , col sep=comma] {./results/tssdn_sync1_sync.csv};
        \addplot[ CoreBlue,
          dotted,
          very thick,
          ] table [x, y , col sep=comma] {./results/tssdn_sync2_sync.csv};
        \addplot[ CoreGreen,
          dashed,
          very thick,
          ] table [x, y , col sep=comma] {./results/tssdn_sync3_sync.csv};
      \nextgroupplot[title={Ordered update},ylabel={End-to-end latency}, y unit=\milli\second]
        \addplot[ CoreRed,
          very thick,
          ] table [x, y , col sep=comma] {./results/tssdn_sync1_ordered.csv};
        \addplot[ CoreBlue,
          dotted,
          very thick,
          ] table [x, y , col sep=comma] {./results/tssdn_sync2_ordered.csv};
        \addplot[ CoreGreen,
          dashed,
          very thick,
          ] table [x, y , col sep=comma] {./results/tssdn_sync3_ordered.csv};
          \legend{Source 1, Source 2, Source 3}
        \nextgroupplot[title={Split ordered update},height=.39\linewidth, ymax=0.00099]
          \addplot[ CoreRed,
            very thick,
            ] table [x, y , col sep=comma] {./results/tssdn_sync1_ordered_fix.csv};
          \addplot[ CoreBlue,
            dotted,
            very thick,
            ] table [x, y , col sep=comma] {./results/tssdn_sync2_ordered_fix.csv};
          \addplot[ CoreGreen,
            dashed,
            very thick,
            ] table [x, y , col sep=comma] {./results/tssdn_sync3_ordered_fix.csv};
            
          \node[small dot,pin={[pin distance=6pt]-100:{C3-1}}] at (0.39,0.00037) {};
          \node[small dot,pin={[pin distance=6pt]-80:{C3-2}}] at (0.4,0.00037) {};
      \end{groupplot}
      
      \node[pin={[pin distance=4pt]90:{C1}}] at (0.608,1.75) {};
      \node[pin={[pin distance=4pt]90:{C2}}] at (1.82,1.75) {};
      \node[pin={[pin distance=4pt]90:{C3}}] at (3.03,1.75) {};
      \node[pin={[pin distance=4pt]90:{C4}}] at (4.242,1.75) {};
      \node[pin={[pin distance=4pt]90:{C5}}] at (5.455,1.75) {};
      \node[pin={[pin distance=4pt]90:{C6}}] at (6.67,1.75) {};
    \end{tikzpicture}
    \captionof{figure}{
      Impact of reconfigurations C1--C6 on end-to-end latency of synchronous traffic for three different update methods.
      Synchronous update -- both switches apply the changes at a synchronized time; 
      ordered update -- changes are executed in the order required by the first operation; 
      split ordered update -- configuration C3 is split in two transactions (C3-1, C3-2). 
    }
    \label{fig:tssdnlatency}
  \end{figure}

\begin{figure*}
  \centering    
  \includegraphics[width=0.8\linewidth,
  trim=0.62cm 0.62cm 0.62cm 0.62cm, clip=true 
  ] {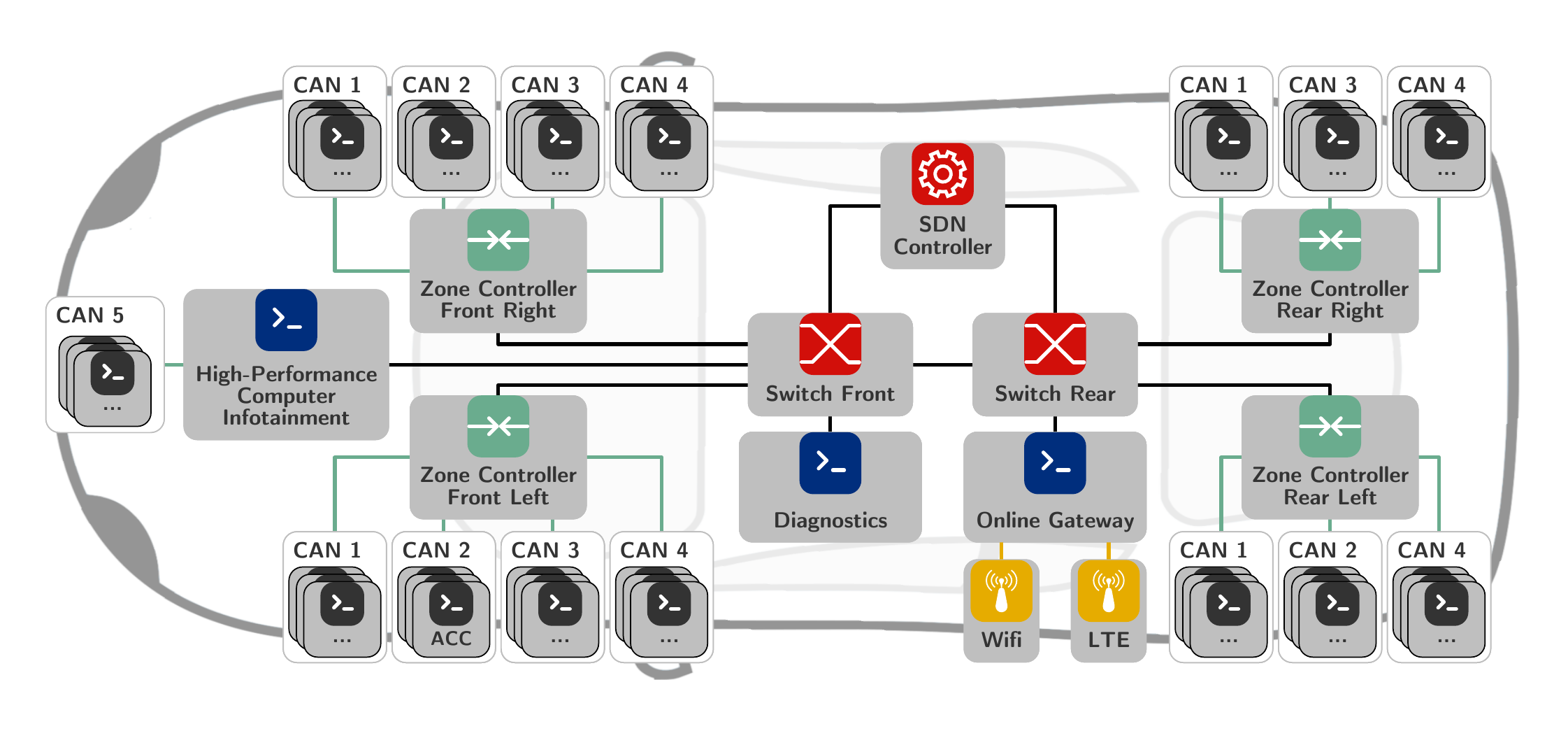}
  \includegraphics[width=0.65\linewidth, trim=0.62cm 0.62cm 0.62cm 0.62cm, clip=true] {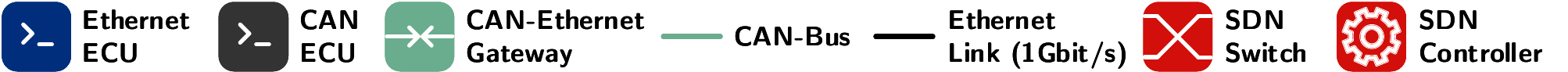}
  \caption{\ac{IVN} of a production vehicle transformed to a software-defined Ethernet network in a zone topology.}
  \label{fig:topo}
  \vspace{-10pt}
\end{figure*}
  
The ordered updates behave identical for all transactions except for C3, since this requires only one commit order. 
The update sequence for C3 is set to change Switch Left first and then Switch Right, which agrees with the commit order for shifting the time slot from S2 to an earlier time and adding a time slot for S3, but conflicts with the commit order for shifting S1 to a later time.
After Switch Left is updated, a frame from S1 must wait additional \SI{175}{\micro\second} according to the new schedule. 
After this frame is transmitted, it misses its time slot at Switch Right, which has not yet been updated.
This delays the frame by \SI{1}{\milli\second} until the next cycle, which can be observed on a few frames of S1 at \SI{0.4}{\second}.
Then Switch Right is updated, which enforces the new schedule but cannot repair the one-cycle delay.
This delays the frame by \SI{1}{\milli\second} until the next cycle, which can be observed on a few frames of S1 at \SI{0.4}{\second}.
Then Switch Right is updated, which enforces the new schedule but cannot repair the one-cycle delay.

Splitting transaction C3 into two transactions (C3-1, C3-2) can solve this problem. 
C3-1 only shifts the flow of S1 first on Switch Right, then on Switch Left.
C3-2 shifts the flow of S2 and adds the flow S4 in reverse order. 
This ensures the correct execution order for all operations in the transaction.

\subsection{Discussion}

Synchronous traffic schedules allow for reconfiguration without a penalty for existing or additional real-time traffic, provided a  suitable update method executes commits simultaneously on all devices or iteratively in the correct order. 
For the latter, the order depends on the type of reconfiguration and requires that a transaction contains only operations of the same update priority.
Synchronous updates can bundle all operations within one transaction.

Controlling asynchronous communication via OpenFlow does not affect real-time traffic, only the \ac{SR} experiences a small start up delay.
Since the \ac{SRP} is transmitted as \ac{BE} traffic, it may be affected by cross traffic, and hence stays without guaranteed temporal bounds in both the \ac{TSSDN} and \ac{TSN}-only variants.
Our additional delay of \SI{0.97}{\ms} is acceptable as it stays well below the automotive requirements for service setup times, which range around \SIrange{150}{200}{\milli\second}. 
To leverage the full potential of \ac{SDN}, the \ac{SRP} can be optimized so that the controller does not need to propagate between switches~\cite{nkm-ssrjr-21}. This could result in an even lower startup delay.

A side effect observed in our evaluations is that \ac{SDN} protects real-time traffic. 
Precise flow matching acts as an ingress control and ensures that new traffic is not forwarded until a flow rule has been installed. 
For transactional schedule reconfiguration, this also implies that the GCL has already been updated prior to accommodating a flow. 
Unknown flows are not queued and thus cannot delay existing real-time traffic.

Finally, our results show that the combination of \ac{SDN} and \ac{TSN} in the proposed architecture works as expected.
Deadlines for synchronous and asynchronous \ac{TSN} flows are met and remain unaffected by the introduction of \ac{SDN}.
Meanwhile, forwarding is controlled by an \ac{SDN} controller, which opens potentials for resilience, security, and adaptability of the \ac{IVN}.

\section{Security Impact of Precise SDN Flow Control for In-Vehicle Communication}
\label{sec:eval_security}

The goal of our security approach with \ac{TSSDN} is to isolate in-vehicle control flows in a shared environment and prevent unwanted traffic. 
We compare the different flow separation concepts described in Section~\ref{subsec:separation} in a case study of a realistic \ac{IVN} that was derived from a production car.

The attacker model for our case-based security study focuses on remote attackers. 
A remote attacker needs access to the car first to launch an attack.  
There are many interfaces, most of which are connected to larger \acp{ECU} such as the infotainment or an online gateway~\cite{cmkas-ceaas-11}.
Once the attacker manages to overcome the first layer of defense, he can gain access to the \ac{IVN} backbone to launch attacks on in-vehicle components.
Correspondingly, our threat model reflects network-centric attacks. Threats arrive from scanning, denial of service, replay, and forgery attacks. We elaborate in characteristic examples, against which threats the \ac{TSSDN} flow separation can safeguard and against which it cannot.

\subsection{Experimental Setup of Our SDN-Based Prototype Car}

\begin{figure}
  \centering
\subfloat[Seat Ateca prototype vehicle\label{fig:prototype}]{%
  \includegraphics[width=0.475\linewidth, 
    height=0.25\linewidth, 
    trim=0.62cm 0.4cm 0.62cm 5.4cm, clip=true]{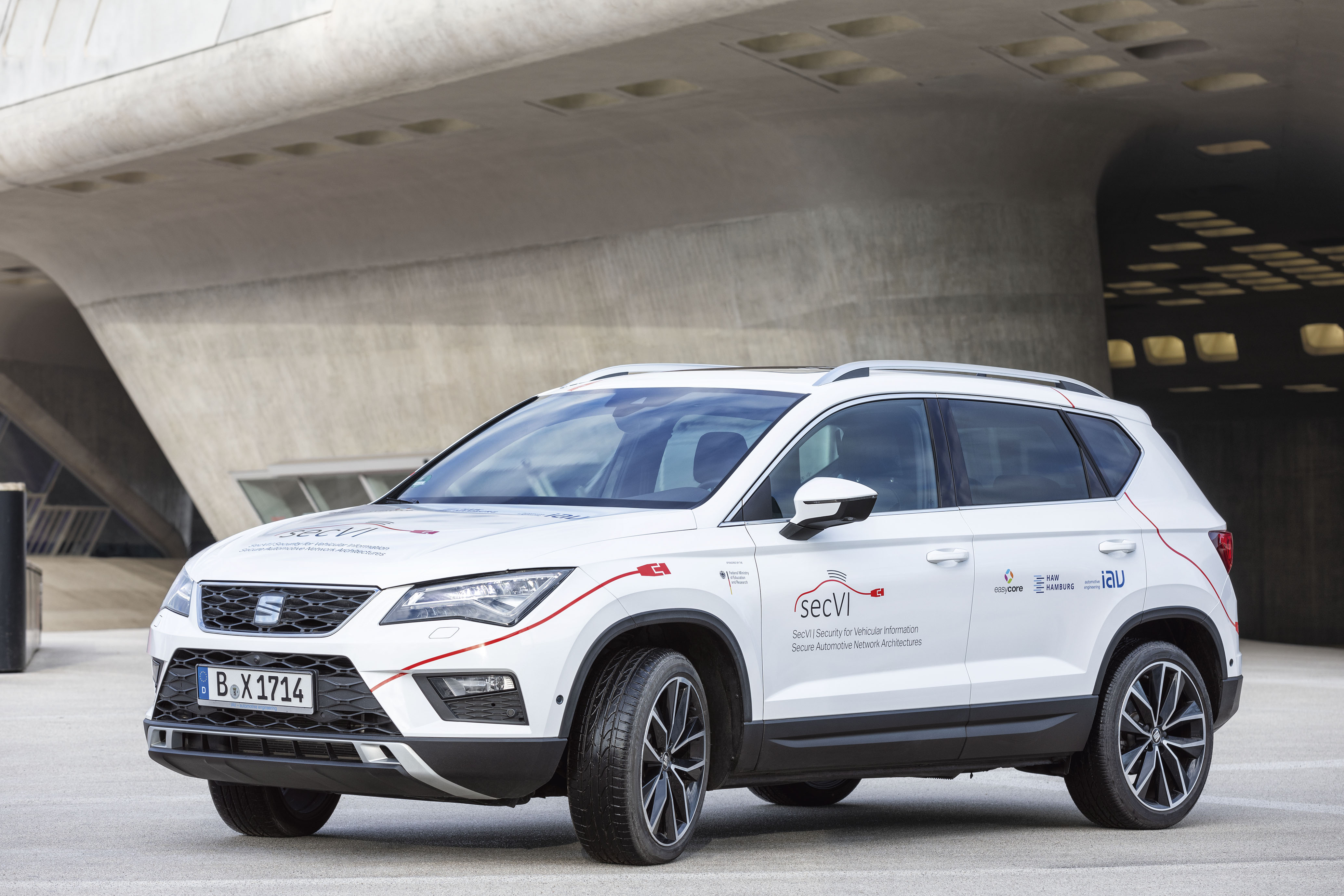}}
  \hfill
\subfloat[Installation in the trunk\label{fig:trunk}]{%
  \includegraphics[width=0.475\linewidth, 
    height=0.25\linewidth, 
    trim=3cm 7cm 3cm 3cm, clip=true]{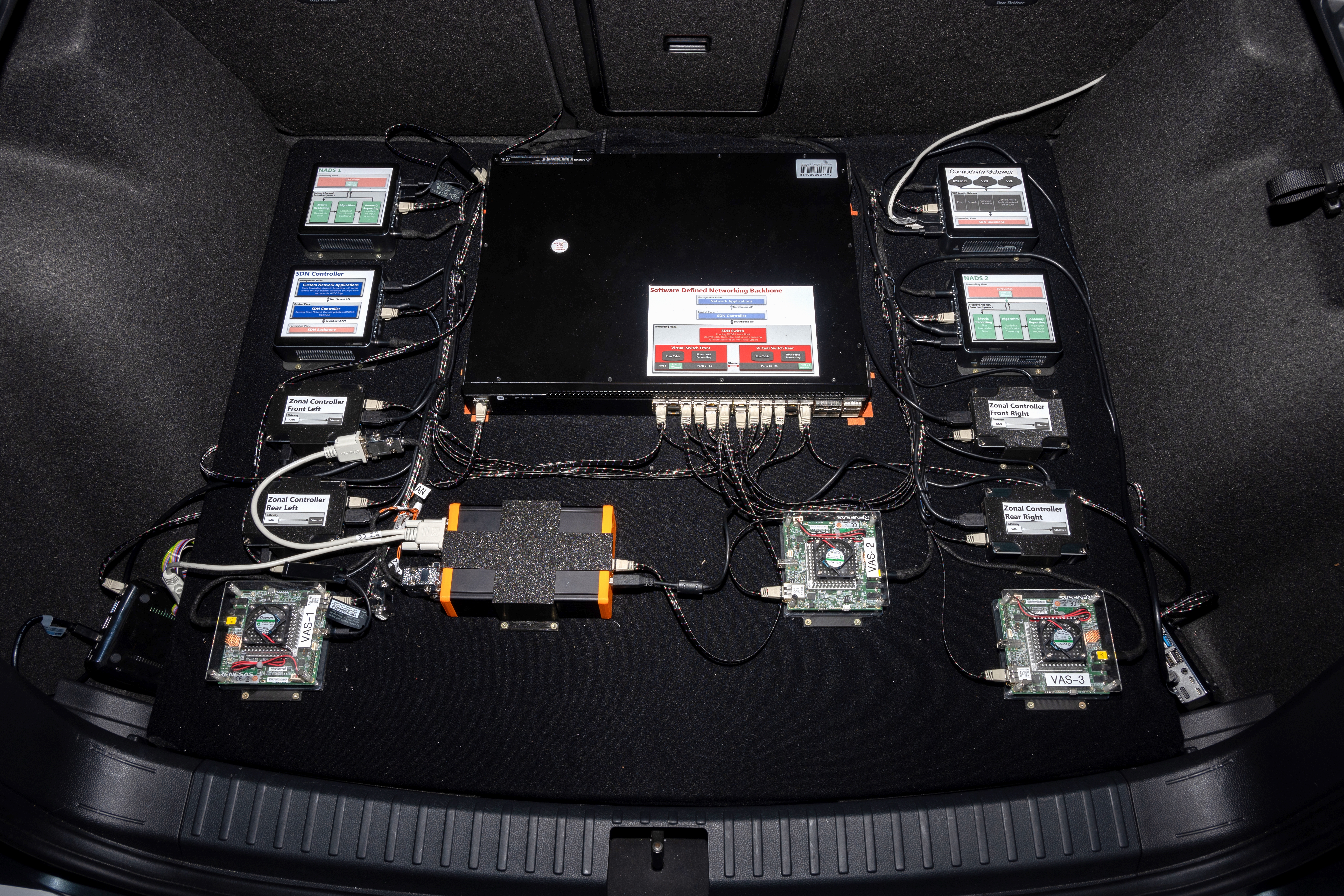}}
\caption{Pictures of the prototype and installed components.}
\label{fig:pictures} 
\vspace{-10pt}
\end{figure}

Figure~\ref{fig:pictures} shows our prototype car (\ref{fig:prototype}) with a software-defined Ethernet backbone installed in the trunk (\ref{fig:trunk}).
The network topology is shown in Figure~\ref{fig:topo} derived from the \ac{CAN} network originally implemented in a central gateway topology, which we transformed into an Ethernet zone topology.
All \ac{CAN} \acp{ECU} are grouped into four zones based on their placement in the vehicle (Front-Left (FL), -Right (FR), Rear-Left (RL), -Right (RR)).
A \acf{ZC} handles computation intensive tasks and acts as a \ac{CAN}-Ethernet gateway.

Despite the placement of the \ac{CAN}-\acp{ECU}, the original domain buses (numbered from 1 to 5) are retained, e.g., CAN 1 on ZC FL and CAN 1 on ZC FR originally belonged to a single domain bus.
A High-Performance Computer handles the infotainment system and acts as a gateway between all CAN \acp{ECU} and the infotainment domain.
We extend the original \ac{CAN} network of the production car with Ethernet \acp{ECU} for diagnostics and external connectivity.

The software-defined Ethernet backbone consisting of two switches controlled by an \ac{SDN} controller.
We generate three network configurations that correspond to our three separation concepts compliant to the communication matrix of our original production vehicle.
In one configuration, \acp{ZC} embed \ac{CAN} messages fully exposed on Layer~2 to enable separation by messages. 
The other two configurations use hidden embedding in SOME/IP tunnels and encode the topic or domain identifier in the multicast destination IP, enabling separation by topic or domain.
The receiving \ac{ZC} transforms the packets back to \ac{CAN} frames and forwards them to the \ac{CAN} bus destinations.

In all three configurations, the original messages from the vehicle are correctly forwarded to all valid receivers.
Our analysis focuses on the invalid control flows that are either blocked or forwarded by the backbone.
The flow separation is not affected by the amount of data transported in the flows and does not depend on the state of the vehicle, e.g., whether it is driving or not. 
In this way, we assume the worst case scenario which allows all original vehicular control flows at all times.
Configuring the network according to the vehicle state can reduce the number of legible flows in the network, which enhances the effect of flow separation, but will not change the characteristic methods of separation.

\subsection{Mapping Control Flows on Network Flows} 
We focus our analysis on the \ac{CAN} Control Flows (CFs) that traverse the Ethernet backbone.
Therefore, we do not consider local CFs, for which the sender and all receivers are located within the same zone.
In total there are 242 different CFs forwarded via the backbone.

\begin{table}[h]
  \centering
  \setlength{\tabcolsep}{5pt}
  \caption{
    Control Flows (CF) bundled in a Network Flow (NF) with the number of NFs that carry multiple CFs.
  }
  \label{tab:controltonetworkflows}
  \begin{tabularx}{\linewidth}{l c c c c}
      \toprule
      \textbf{Separation} & \multirow{2}{2cm}{\textbf{\# NFs (with multiple CFs)}} 
      & \multicolumn{3}{c}{\textbf{\# CFs per NF}}
      \\
      & & Minimum & Average & Maximum       
      \\ \midrule
      \textbf{By Message} & 242 \phantom{0}(0) & 1 & \phantom{0}1 & \phantom{0}1
      \\
      \textbf{By Domain} & \phantom{0}19 (19) & 5 & 13 & 37 
      \\
      \textbf{By Topic} & 102 (38) & 1 & \phantom{0}3 & 17
      \\ 
      \bottomrule
  \end{tabularx}
\end{table}

Table~\ref{tab:controltonetworkflows} shows the generated Network Flows (NFs) for each separation concept in relation to the number of CFs bundled in a NF.
Message separation isolates the 242 CFs within individual NFs.
Separation by domain generates 19 NFs, one tunnel for each sender and domain all of which carry multiple CFs.
There is a minimum of 5, an average of 13, and a maximum of 37 CFs per NF, so there is at least one NF carrying 37 CFs that cannot be distinguished by the network.
The topic separation generates 102 NFs, one tunnel for each sender and topic.
38 of these NFs carry multiple CFs.
This means that 64 of the topics contain only one CF and therefore behave exactly as if separated by message.
In the following analysis, separation by message can be used as a baseline as it implements exactly the relations of the communication matrix.

\begin{figure}
	\centering
    \begin{tikzpicture}
        \begin{axis}
          [width=\linewidth,height=.5\linewidth,
          ybar,
          bar width=9pt,
          legend cell align={left},
          legend style={anchor=north east, font=\small},
          legend columns=-1,
          xtick={1,...,4},
          xtick pos=bottom,
          xticklabels={1, 2, 3, 4},
          x tick label style={align=center},
          xmin=1,
          xmax=4,
          enlarge x limits=0.2,
          xlabel={Destination ZCs [\#]},
          ylabel={Network flows [\%]}, 
          ytick = {0,20,40,60,80,100}, 
          ymin = 0,
          ymax = 105,
        ]
            \addplot[fill=CoreRed] 
            coordinates { (1,27.27) (2,16.12) (3,45.45) (4,11.16)};
            \addplot[fill=CoreYellow] 
            coordinates { (1,25.49) (2,15.69) (3,41.18) (4,17.65)};
            \addplot[fill=CoreGreen] 
            coordinates { (1,0) (2,0) (3,31.58) (4,68.42)};
            \legend{Separation by Message, by Topic, by Domain}
        \end{axis}
    \end{tikzpicture}
    \vspace{-4pt}
    \caption{
      Destination ZCs reached by a share of NFs.
      }
    \label{fig:cfreachinggateways}
    \vspace{-10pt}
\end{figure}
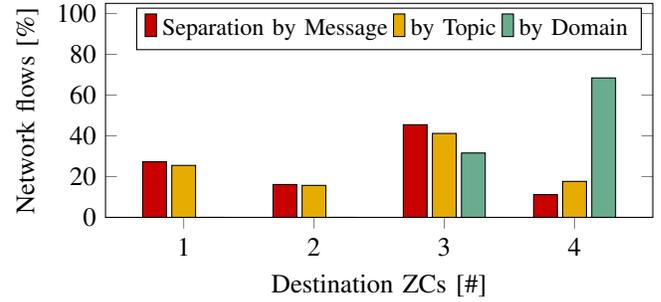 

All NFs are implemented as multicast to reach multiple destinations if necessary.
Figure~\ref{fig:cfreachinggateways} shows the share of NFs sent to a particular number of destinations (ZCs).
Separation by message and topic show similar results, although additional 7\% of NFs reach all gateways with separation by topic.
Again, this similarity could be caused by the 64 CFs in completely isolated topics.
With domain separation zero NFs reach fewer than three gateways, around a third reach exactly three and nearly 70\% reach all gateways in the network.
In particular, the relative difference of NFs reaching all four destinations is notable.
This is expected as the \acp{ECU} of a domain are usually distributed throughout the vehicle and therefore most domains are present in every zone.
Nevertheless, this is a security issue as those receivers should not be able to receive those flows and also indicates that the embedding has a big influence on network overhead. 

\subsection{Security and Performance of Control Flow Separation}
The attack surface of CFs depends on their isolation in the network.
We compare the separation concepts based on how well CFs can be attached to paths between the ZCs.
Without separation, any node in the network could send and receive any of the 242 different CFs traversing the backbone.
Optimally, however, only gateways that actually link \acp{ECU} participating in a CF should be able to send and receive it on the backbone.
To evaluate this, we sort the 242 possible CFs that could be sent from one ZC to another into the following categories:
\begin{itemize}
  \item A CF is \textbf{legitimate} if the original CAN source \ac{ECU} is connected to the source ZC and at least one CAN \ac{ECU} connected to the destination ZC is a valid receiver according to the communication matrix.
  \item An \textbf{oversupplied} CF is legitimately sent but received by a destination ZC that does not need it. 
  \item A CF is \textbf{permitted} if it could be sent by the source and would be forwarded to the destination. 
  Even though not present in our communication matrix, these flows could be used by malicious components.
  \item A \textbf{forbidden} CF is filtered by the backbone and is not forwarded from the source to the destination.
\end{itemize}

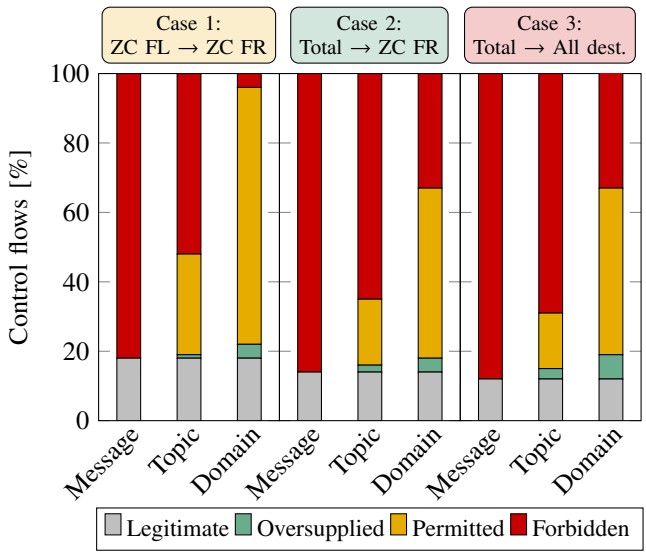
\begin{figure}
  \centering
  \begin{tikzpicture}
    \begin{groupplot}[
      group style={
        group size=3 by 1, 
        horizontal sep=0pt, 
        vertical sep=0pt,  
        yticklabels at=edge left, 
      },
      width=.45\linewidth,
      height=.7\linewidth,
      legend style={at={(-0.5,-0.25)}, anchor=north, legend columns=-1, font=\small},
      ymin=0, 
      ymax=100,
      xticklabel style={anchor=north east, yshift=4pt, xshift=0pt,rotate=45},
      xtick pos=bottom, 
      symbolic x coords ={Message, Topic, Domain},
      xtick=data, 
      enlarge x limits=0.25,
    ]

      \nextgroupplot[ybar stacked, bar width=9pt, ylabel={Control flows [\%]},]
        \addplot[fill=CoreGray] coordinates {(Message,18) (Topic,18) (Domain,18)};
        \addplot[fill=CoreGreen] coordinates {(Message,0) (Topic,1) (Domain,4)};
        \addplot[fill=CoreYellow] coordinates {(Message,0) (Topic,29) (Domain,74)};
        \addplot[fill=CoreRed] coordinates {(Message,82) (Topic,52) (Domain,4)};
      
      \nextgroupplot[ybar stacked, bar width=9pt,]
        \addplot[fill=CoreGray] coordinates {(Message,14) (Topic,14) (Domain,14)};
        \addplot[fill=CoreGreen] coordinates {(Message,0) (Topic,2) (Domain,4)};
        \addplot[fill=CoreYellow] coordinates {(Message,0) (Topic,19) (Domain,49)};
        \addplot[fill=CoreRed] coordinates {(Message,86) (Topic,65) (Domain,33)};
      
        \nextgroupplot[ybar stacked, bar width=9pt,]
          \addplot[fill=CoreGray] coordinates {(Message,12) (Topic,12) (Domain,12)};
          \addplot[fill=CoreGreen] coordinates {(Message,0) (Topic,3) (Domain,7)};
          \addplot[fill=CoreYellow] coordinates {(Message,0) (Topic,16) (Domain,48)};
          \addplot[fill=CoreRed] coordinates {(Message,88) (Topic,69) (Domain,34)};
          
      \legend{
        Legitimate, 
        Oversupplied, 
        Permitted, 
        Forbidden,
      }
    \end{groupplot}
    \tikzstyle{every node}=[font=\footnotesize]
    \node[draw,shape=rectangle, rounded corners=3pt,above, black, fill=LightCoreYellow, align=center, ] at (1.2,4.75) {Case 1: \\ZC FL $\rightarrow$ ZC FR};

    \node[draw,shape=rectangle, rounded corners=3pt,above, black, fill=LightCoreGreen, align=center] at (3.6,4.75) {Case 2: \\Total $\rightarrow$ ZC FR};

    \node[draw,shape=rectangle, rounded corners=3pt,above, black, fill=LightCoreRed, align=center] at (6.0,4.75) {Case 3: \\Total $\rightarrow$ All dest.};
  \end{tikzpicture}
  \caption{Share of legitimate, oversupplied, permitted, and forbidden CFs from ZC FL to ZC FR, all sources to ZC FR, and all sources to all destinations.
  }
  \label{fig:oversupply}
  \vspace{-10pt}
\end{figure}

Figure~\ref{fig:oversupply} visualizes the shares of legitimate, oversupplied, permitted and forbidden CFs in relation to the maximum possible CFs in three cases: 
(1) concentrates on a specific connection between two exemplary ZCs (from ZC FL to ZC FR), 
(2) includes all CFs from any source ZC to one destination (ZC FR), 
and (3) looks at the whole backbone communication from all ZCs to all ZCs.
An exhaustive analysis of all communication relations between the gateways in our prototype can be found in~\cite{hsmks-sicfs-20}.

All three cases clearly show the effectiveness of the strict message separation from exposed embedding.
When separated by message, the software-defined backbone only forwards the legitimate CFs and forbids any other CF.
The hidden application layer embeddings oversupply and permit illegitimate CFs.
Still, only CFs of the tunnels that are already in use by the sending nodes are supported by the network.
More precisely, topics can improve the separation of flows in the network.
As explained earlier, we chose our topics by grouping similar messages from the communication matrix. 
We could not group some CFs into topics, making them perfectly separated in their own topic, which affects the results.

In the third case, 12\% of all CFs from all sources to all destinations are legitimate.
With separation by topic additional 3\% are oversupplied at the destinations, with domain separation around 7\%.
Gateways can filter the unneeded CFs at the cost of processing power so that no unwanted message reaches a \ac{CAN} bus.
Still, the network load on the backbone increases through oversupplied control traffic with domain and topic separation. 
This clearly conflicts with the benefits of aggregation, which is one of the main arguments for using hidden embeddings.

\begin{table}
    \centering
    \caption{
        Sent and received bandwidth at all four ZCs of Ethernet embedded CAN traffic (CAN Bus 1 -- 4) separated by message or domain with different aggregation intervals.
    }
    \label{tab:aggregation}
    \setlength{\tabcolsep}{4pt}
    \begin{tabularx}{\linewidth}{p{1.5cm} C{1.5cm} C{1.4cm} C{1.45cm} C{1.45cm} }
        \toprule
        \textbf{Embedding} & \textbf{Aggregation interval} & \textbf{Average frame size} & \textbf{Bandwidth sent} & \textbf{Bandwidth received} \\
        \midrule
        Per message & w/o agg. & \phantom{0}\SI{64}{\byte} & \SI{1.4}{\mega\bit\per\second} & \SI{3.1}{\mega\bit\per\second}\\
        \cmidrule(rl){1-5}
        \multirow{6}{*}{\parbox{1.45cm}{Per domain}} & w/o agg. & \phantom{0}\SI{72}{\byte} & \SI{1.5}{\mega\bit\per\second} & \SI{4.6}{\mega\bit\per\second}\\
         & \phantom{00}\SI{1}{\milli\second} & \SI{101}{\byte} & \SI{1.1}{\mega\bit\per\second} & \SI{3.3}{\mega\bit\per\second}\\
         & \phantom{00}\SI{5}{\milli\second} & \SI{119}{\byte} & \SI{1.0}{\mega\bit\per\second} & \SI{2.9}{\mega\bit\per\second}\\
         & \phantom{0}\SI{10}{\milli\second} & \SI{145}{\byte} & \SI{0.9}{\mega\bit\per\second} & \SI{2.6}{\mega\bit\per\second}\\
         & \phantom{0}\SI{50}{\milli\second} & \SI{324}{\byte} & \SI{0.7}{\mega\bit\per\second} & \SI{2.1}{\mega\bit\per\second}\\
         & \SI{100}{\milli\second} & \SI{527}{\byte} & \SI{0.6}{\mega\bit\per\second} & \SI{1.9}{\mega\bit\per\second}\\
        \bottomrule
    \end{tabularx}
\end{table}

Table~\ref{tab:aggregation} shows the average bandwidth sent and received by all four ZCs and the frame size of Ethernet-embedded CAN traffic from our production vehicle over a period of \SI{60}{\second}. 
With exposed and hidden embedding, about the same bandwidth is sent without aggregation.
The average frame size increases when multiple CAN frames are aggregated for a given interval before being sent alone or together with other messages, which in turn significantly increases latency and jitter.
This is not possible for exposed embeddings, since only one message per frame is allowed.
Most evaluations of aggregation strategies only consider the bandwidth sent by the gateways. 
Our evaluation shows that the embedding strategy and thus the enabled separation in the network has a major impact on the bandwidth at the receiver side.
Without aggregation, separation by domain results in 50\% higher received bandwidth, again illustrating oversupply.
The transmitted and received bandwidth depends on the traffic pattern and network topology.
Kern et al.~\cite{krst-gsece-11} were able to reduce the transmitted bandwidth by up to 75\% using synthetic CAN traffic when messages are aggregated for \SIrange{5}{10}{\milli\second}. 
For our specific vehicle, aggregation for an interval of \SI{10}{\milli\second} reduces the bandwidth sent by 35\% while the received bandwidth is reduced by only 6\%. 
The evaluation shows that aggregating traffic from low-bandwidth bus systems such as CAN has little impact, especially when considering \SI{1}{\giga\bit\per\second} Ethernet networks.
Above all, it is a security issue that network nodes are confronted with messages for which they are not prepared. 
Separation by message is the only concept that can solve this problem  with zero oversupplies.

From a security perspective, the CFs that a (malicious) \ac{ECU} could send are even more relevant than the CFs it can receive.
From all sources to all destinations, 31\% of all CFs are permitted with topic separation and even 67\% with separation by domain, while only 12\% are needed (see Figure~\ref{fig:oversupply}).
This shows the security weakness of the hidden embedding approach.
When a device participates in a tunnel, it can listen to all CFs of this tunnel and is permitted to send all CFs of the tunnel.
A smaller number of filtered CFs eases attacks as fewer \acp{ECU} need to be compromised to gain control of the car. 
Even if gateways filter all unwanted and illegally sent messages it would still be possible to attack the NFs, e.g., by flooding the tunnels to delay time-critical flows.
In the case of exposed embeddings, which fully isolates messages by SDN, attackers must compromise the exact sender of each CF on the Ethernet backbone to issue messages in this channel.

\subsection{Benefits and Limitations in Characteristic Attack Scenarios}
\label{subsec:attack_scenarios}
We investigate how robustly our setup of static flows augmented by ACL-controlled  dynamic flows can isolate in-vehicle control flows and prevent unwanted traffic in case of example attacks.
The attack surface of the \ac{IVN} depends heavily on which data flows are blocked or forwarded. 
We consider examples that affect data flows in the network to illustrate the impact of SDN flow control.

For characteristic attack scenarios we assume that the attacker succeeded in compromising the gateway so that attacks can be launched from there.
Table~\ref{tab:generic_attacks} summarizes the results for our \ac{IVN} with and without SDN access control. 
For the latter, the controller acts as a simple Ethernet switch. 

\begin{table}[ht]
  \caption{Attacks launched from the online gateway in an unprotected environment and with SDN access control. A host scan searches for the existing 11 hosts, a port scan examines 4 open TCP and 8 open UDP ports on a host, and a TCP SYN flood of 1000 connections tries to impair the target. }
  \label{tab:generic_attacks}
  \setlength{\tabcolsep}{4pt}
  \begin{tabularx}{\linewidth}{l l l }
    \toprule
    \textbf{Attack impact} & \textbf{w/o access control} & \textbf{w/ access control}
    \\
    \midrule
    Hosts discovered & 11 & 11 (0 when arp is blocked) \\
    Ports discovered & 4 TCP, 8 UDP & 0 TCP, 0 UDP \\
    TCP SYN flood arrived & 1000 & 0 (1000 at SDN controller) \\
    \bottomrule
  \end{tabularx}
\end{table}

Scans are the most common network attacks and used to gather information about the system.
All eleven hosts can be discovered from the online gateway with \textit{nmap} even with access control as long as ARP is not blocked by the ACL. 
Since hosts in the \ac{IVN} are usually known, ARP can be blocked, which means that no hosts are discoverable.
Scanning the ports of a host without access control, we discover 4 open TCP ports and 8 open UDP ports.
With access control, no open ports can be discovered.
This is true even for ports to which there is a connection, since the flow rules match source and destination ports and \textit{nmap} uses the wrong source port.
With elevated rights, however, the attacker could select the correct source port, which requires detailed knowledge about the vehicle and is even more difficult on embedded devices.
Without scans, most attackers cannot perform further attacks.

As an example of a denial of service attack, we perform a TCP SYN flood attack on one of the discovered TCP ports.
Without access control, all 1000 SYNs arrive at the target and the attack is successful.
Since new source port and IP pairs are used for each SYN, they are detected and blocked by our \ac{ACL}.
Such attacks, however, are then forwarded by the switches to the SDN controller and could overload it.
This is a known problem in \ac{SDN} and protection mechanisms have been investigated in the past~\cite{hjtuj-cssjr-20}.
For example, most current SDN controllers can be distributed across multiple instances in standby to eliminate the single point of failure. 
In our case, the static flow rules in the switches are preserved even when the controller is unavailable, which guarantees safety-critical communication.
Nevertheless, protection mechanisms for the controller should be further investigated in future work.

\begin{table}
  \caption{Share of packets forwarded from a 30s replay (10206 packets total) recorded at ZC FL and played back from the online gateway.
  The results are shown for exposed and hidden embeddings each with SDN access control and conventional L2/L3 multicast. 
  For the conventional forwarding unknown multicast groups are either dropped or broadcasted. 
  }
  \label{tab:replay_attack}
  \setlength{\tabcolsep}{2pt}
  \begin{tabularx}{\linewidth}{l l c c c }
    \toprule
    \textbf{Embedding} & \textbf{Dest.} & \multicolumn{2}{c}{\textbf{Conventional multicast}} & \textbf{w/ access control} 
    \\
    && (drop policy) & (broadcast policy) & 
    \\
    \midrule
    \multirow{4}{*}{\parbox{0.13\linewidth}{Per \\domain \\(hidden)}}
    &ZC FL & 10206 (100\%) & 10206 (100\%) & 0
    \\
    &ZC FR & 10206 (100\%) & 10206 (100\%) & 0
    \\
    &ZC RR & 10206 (100\%) & 10206 (100\%) & 0
    \\
    &ZC RL & 10206 (100\%) & 10206 (100\%) & 0
    \\
    \cmidrule(rl){1-5}
    \multirow{4}{*}{\parbox{0.13\linewidth}{Per \\message \\(exposed)}}
    &ZC FL & \phantom{0000}0\phantom{ (000\%)} & \phantom{0}1903 \phantom{0}(19\%) & 0 
    \\
    &ZC FR & \phantom{0}7242 \phantom{0}(71\%) & 10022 \phantom{0}(98\%) & 0
    \\
    &ZC RR & \phantom{0}8617 \phantom{0}(84\%) & \phantom{0}9525 \phantom{0}(93\%) & 0
    \\
    &ZC RL & \phantom{0}4072 \phantom{0}(40\%) & \phantom{0}4980 \phantom{0}(49\%) & 0
    \\
    \bottomrule
  \end{tabularx}
\end{table}

For targeted attacks, we assume that the adversary has detailed knowledge about the vehicle and attacks specific components.
We analyze the impact of a replay attack, which is a targeted attack on multiple control flows.
We recorded a \SI{30}{\second} trace of 10206 packets at ZC FL embedded exposed and hidden (per domain).
Table~\ref{tab:replay_attack} shows the results for a replay from the online gateway.
For conventional multicast, network device can react to unknown multicast addresses by either dropping or broadcasting the packet.
Multicast groups that reach only devices on one switch are not registered on the other switch. 
With domain embedding, both switches are aware of all multicast groups, so all packets reach all destinations with both policies.
Using per message embedding performs better because not every device joins all multicast groups.
The original sender (ZC FL) does not receive packets with the drop policy because it is not a receiver of any of the multicast groups.
The valid receivers, however, receive all packets from their registered multicast groups.
With SDN access control, all packets are blocked since the ingress port connected to the online gateway is not a valid source of any of the flows. 
Herein lies the strength of our network-centric security approach. 
Regardless of the embedding, an attacker must control a legitimate sender of a CF in order to send it. 

In our \ac{IVN}, there are no flows from the online gateway to the zonal controllers, so no direct driving commands can be sent.
In future autonomous scenarios, larger \acp{ECU} will also use online services, e.g., for detailed up-to-date maps, and thus have communication paths to the online gateway.
Targeted attacks, such as a, replay, packet flood, or forged information in an established network flow, are not prevented by our approach. 
Additional mechanisms such as anomaly detection can help detecting such attacks and countermeasures could be initiated by \ac{SDN}, e.g., through reconfigurations.

\subsection{Discussion} 
Our evaluation shows how \ac{SDN} flow control can protect the \ac{IVN}.
In general, communication is drastically limited and attacks on unknown flows are detected and blocked.
Regardless of the embedding strategy, Ethernet \acp{ECU} such as the online gateway are not allowed to send or receive control messages, which reduces the attack surface.
This protects legacy \acp{ECU} that lack defense mechanisms from attacks by stronger \acp{ECU} as communication is forbidden between them.

On the other hand, targeted attacks that use flows already installed in forwarding devices cannot be prevented by our access control mechanism.
Hidden embeddings lead to unintended receivers of critical CFs and permit their transmission from other network participants, posing a risk to safety and security.
Exposed embedding allows the network to separate all CFs, ensuring that only the original sending \ac{ZC} is allowed to send the CF and that it only reaches the necessary destinations.
This adds a layer of network security even in the case of a compromised gateway, but cannot prevent attacks from \acp{ECU} that are allowed to send certain CFs.
Nevertheless, attacks are limited to the pre-installed and allowed flows and thus the attack surface is significantly reduced.

Our zone topology with SDN-enforced message separation is also more secure than the original network architecture of our production vehicle.
In a pure CAN bus architecture, traffic on the buses cannot be controlled.
All \acp{ECU} connected to a domain bus can send any CF and receive any message on the bus.
A compromised \ac{ECU} can attack all other \acp{ECU} in its domain.
Gateways can use the CAN IDs to filter messages to be forwarded, but cannot verify the correct sender of a CF.
The zone topology splits the CAN domain buses so that fewer \acp{ECU} are connected to a physical bus.
The gateways can thus filter messages and fewer \acp{ECU} are completely unprotected.
If each \ac{ECU} were directly connected to the zone controller, the correct sender could be verified.
Combined with separation of messages by SDN, this could offer perfect separation of CFs and thus a robust, trustworthy communications backbone.

In all attack scenarios, the SDN controller must be well protected because it could be the target of attacks itself.
In our case, the static configuration protects the safety-critical traffic in the \ac{IVN}, since it cannot be changed by the controller.
Future work could perform a risk analysis for attacks on the SDN controller in vehicles.

\section{Conclusion and Outlook}
\label{sec:conclusion_outlook}
In this work, we investigated the integration of \ac{TSN} with \ac{SDN} for improving network security in Ethernet-based \acp{IVN}.
We presented a \ac{TSSDN} switching architecture that harmonizes the functions of \ac{TSN} and \ac{SDN}. 
At its core, our approach implements \ac{SDN} flow control for simultaneous asynchronous and synchronous real-time as well as best-effort traffic in vehicles.
We could show how time-sensitive flows can be reserved via OpenFlow, and how \ac{TDMA} schedules can be reconfigured at runtime without sacrificing the real-time capabilities of \ac{TSN}.

Targeting at network-level security by isolation, we comparatively evaluated three strategies for mapping control flows into a software-defined Ethernet backbone of a real-world \ac{IVN}, which we transformed into a realistic software-defined Ethernet topology. 
Our analysis revealed that network security and performance can be largely improved by exposing control flow properties in standard network header fields which are processed by forwarding devices. 
Embeddings that are hidden within application layer protocols lead to significant oversupply of control flows, which opens the attack surface. 

Future work shall analyze the impact of network-wide schedule reconfiguration strategies on real-time traffic in different scenarios.
A mechanism for TSN senders of scheduled traffic is needed to exchange requirements with the SDN controller and announce the start of a transmission. 
Additional network-level intelligence can further improve in-vehicle security, as there are still unused options applicable as security guards based on \ac{SDN} monitoring and control.


%




\ifCLASSOPTIONcaptionsoff
  \newpage
\fi



\bibliographystyle{IEEEtran}

\bibliography{HTML-Export/all_generated,rfcs,local}
%



\begin{IEEEbiography}
  [{\includegraphics[width=1in,height=1.25in,clip,keepaspectratio]{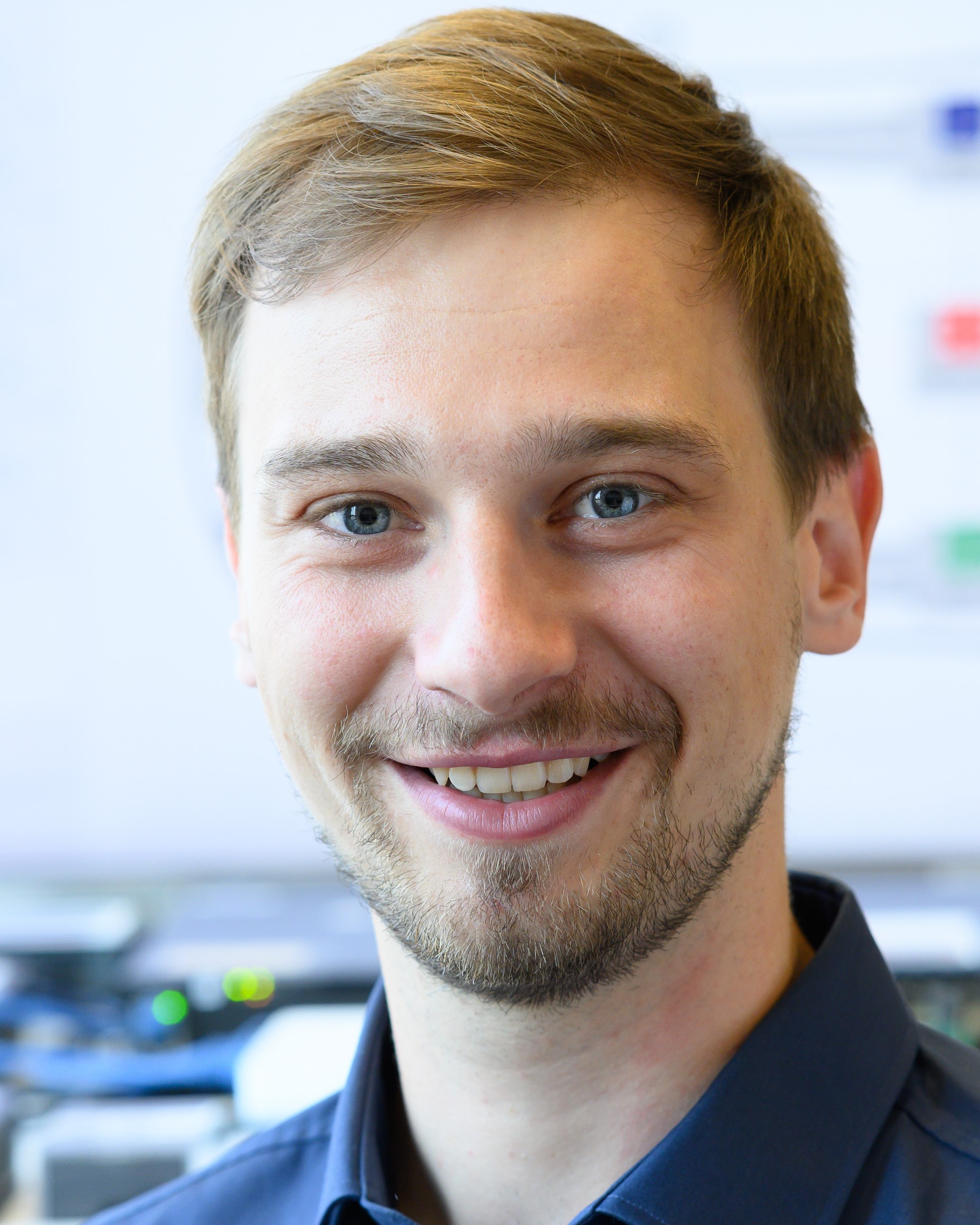}}]
  {Timo H\"ackel}
  received his M.Sc. degree in computer science from the Hamburg University of Applied Sciences (HAW), Hamburg, Germany, in 2018, where he is currently pursuing the Ph.D. degree with the Internet Technologies Research Group.
  His particular research interest is the security of time-sensitive vehicular networks, which he is exploring within the German research project SecVI -- Security for Vehicular Information.
  Timo H\"ackel is also part of the Communication over Real-Time Ethernet research group at HAW Hamburg, where he develops and maintains open source frameworks for the \mbox{OMNeT++} simulator including the CoRE4INET and SDN4CoRE framework.
\end{IEEEbiography}

\vspace*{-2\baselineskip}
\begin{IEEEbiography}[{\includegraphics[width=1in,height=1.25in,clip,keepaspectratio]{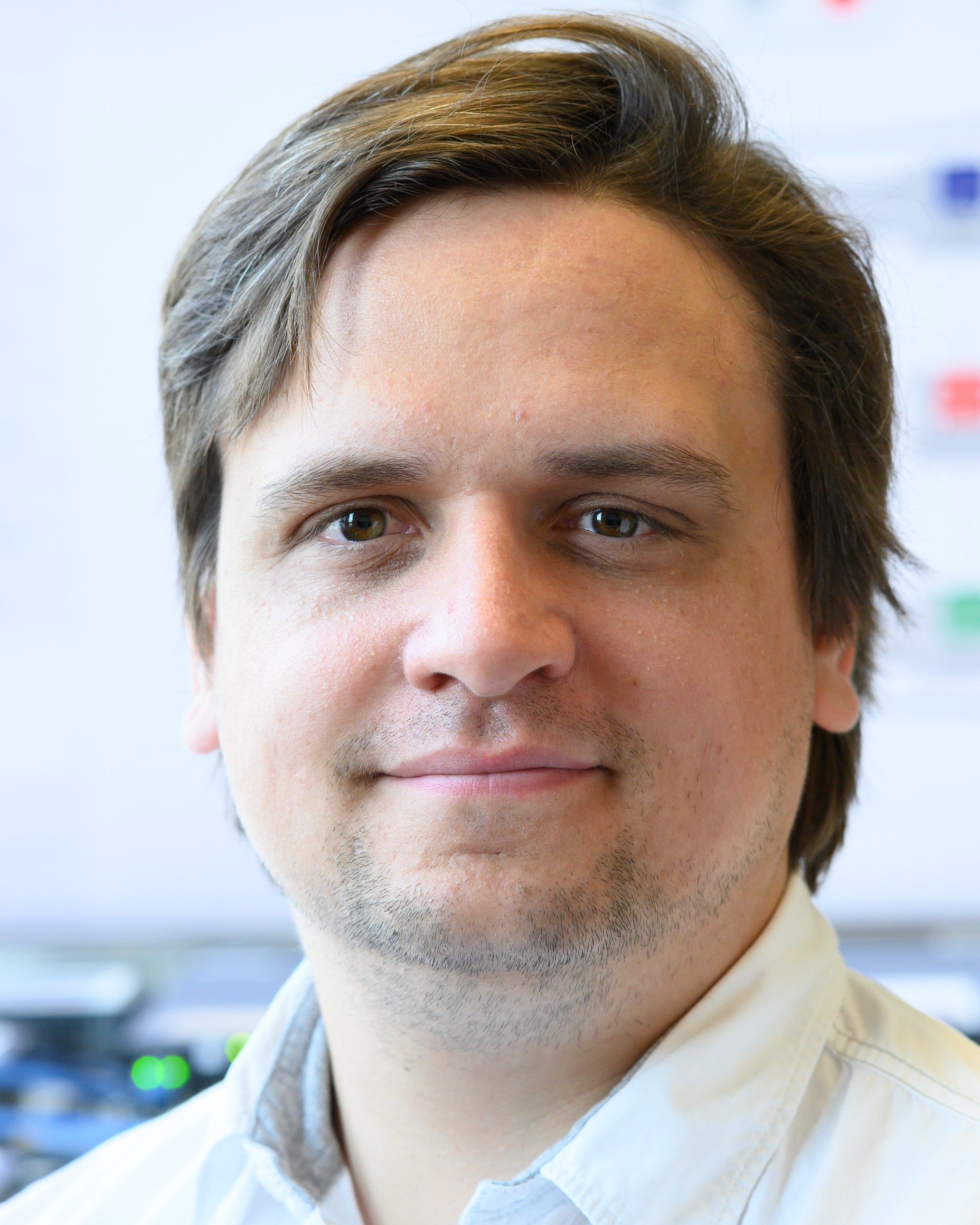}}]
  {Philipp Meyer} is Ph.D. student and research assistant in the Communication over Real-Time Ethernet (CoRE) Research Group at the Hamburg University of Applied Sciences (HAW), Hamburg, Germany. As part of the CoRE Research Group, he started 2012 with researching on real-time Ethernet technologies and their security. He received his B.Sc. (2013) and M.Sc. (2018) in computer science in this context. Furthermore, he pursued his interests as part of the research projects Realtime Ethernet Backbone for Cars (RECBAR) and X-Check. Currently, he is exploring in-car communication security as part of the research project Security for Vehicular Information (SecVI). Philipp also develops and maintains the CoRE open-source simulation environment and frameworks for  \mbox{OMNeT++}.
\end{IEEEbiography}

\vspace*{-2\baselineskip}
\begin{IEEEbiography}[{\includegraphics[width=1in,height=1.25in,clip,keepaspectratio]{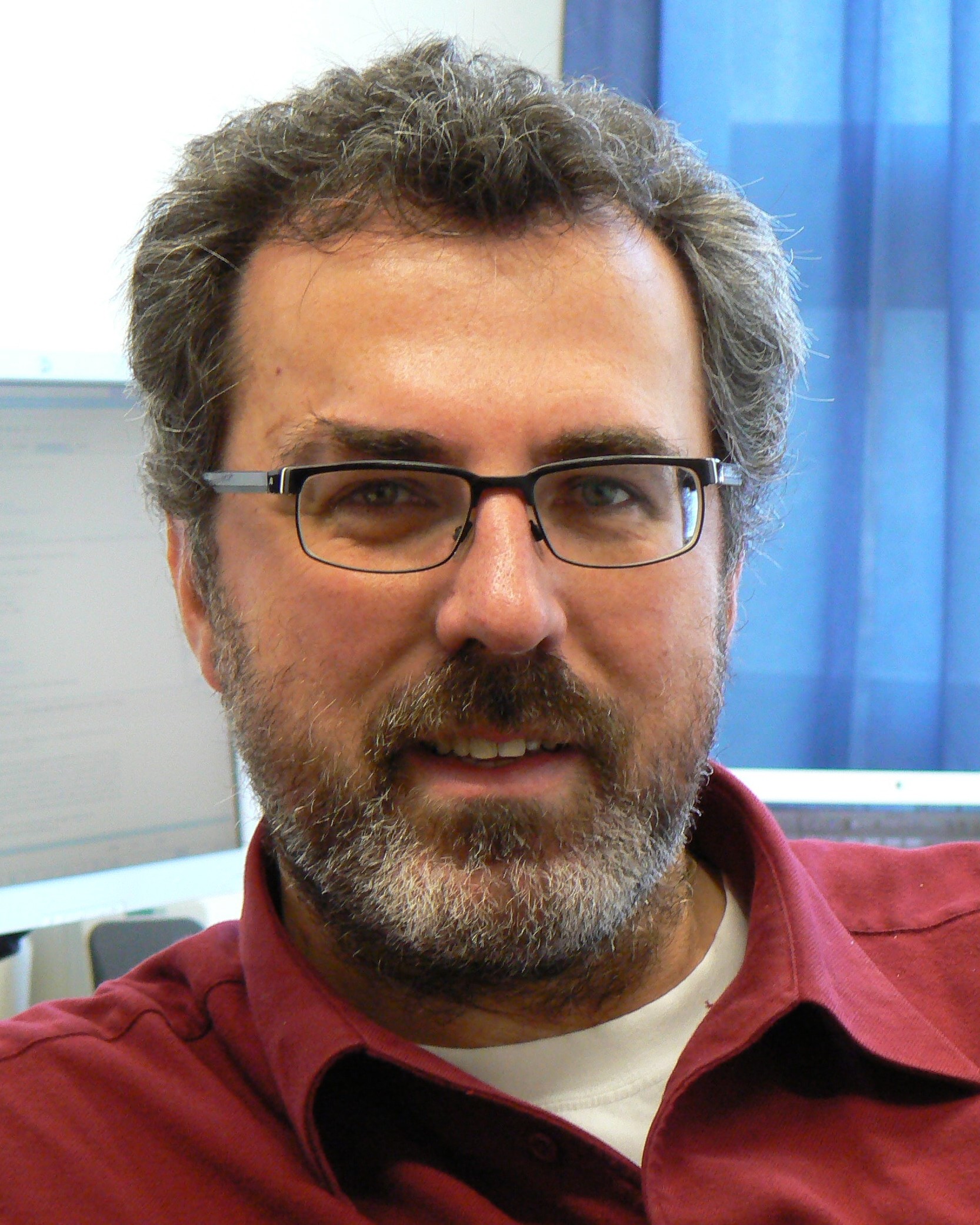}}]
	{Franz Korf} is professor of Embedded Systems at Hamburg University of Applied Sciences (HAW), where he heads the Communication over Real-Time Ethernet research group. He studied computer science at RWTH Aachen University. At the chair of W. Damm at the University of Oldenburg, he received his doctorate in the field of system-level synthesis tools. Before joining HAW Hamburg in 2004, he headed the OEM development of server systems at Fujitsu Siemens Computers. At HAW Hamburg, Franz was responsible for various R \& D projects in the areas of real-time Ethernet architectures and embedded systems.
\end{IEEEbiography}

\vspace*{-2\baselineskip}
\begin{IEEEbiography}[{\includegraphics[width=1in,height=1.25in,clip,keepaspectratio]{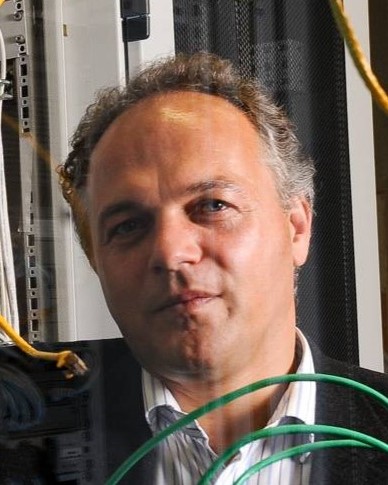}}]
  {Thomas C. Schmidt}
is professor of Computer Networks and Internet Technologies
at Hamburg University of Applied Sciences (HAW), where he heads the Internet
Technologies research group (iNET). 
  He studied mathematics, physics and German literature at Freie Universitaet Berlin and University of Maryland, and received his Ph.D. from FU Berlin in 1993. Since then he has continuously conducted numerous national and international research projects. His continued interests lie in the development, measurement, and analysis of large-scale distributed systems like the Internet. He serves as co-editor and technical expert in many occasions and is actively involved in the work of IETF and IRTF.  Thomas is a co-founder of several large open source projects and coordinator of the community developing RIOT---the friendly Operating System for the IoT.
\end{IEEEbiography}

\end{document}